\documentclass[trackchanges, twocolumn]{aastex701}
\usepackage[flushleft]{threeparttable}

\usepackage{amsmath}	
\usepackage{caption}
\usepackage{subfig}

\begin{document}

\title{PAHSPECS: Spatially Resolved PAH Spectroscopy at cosmic noon with JWST MIRI MRS }

\author[orcid=0000-0002-6460-3682,sname='Donnan']{Fergus R. Donnan}
\affiliation{Department of Astrophysics, University of California San Diego, 9500 Gilman Drive, San Diego, CA 92093, USA}
\email[show]{fdonnan@ucsd.edu}  

\author[orcid=0000-0002-4378-8534,sname='Sandstrom']{Karin Sandstrom}
\affiliation{Department of Astrophysics, University of California San Diego, 9500 Gilman Drive, San Diego, CA 92093, USA}
\email[]{kmsandstrom@ucsd.edu}  

\author[0000-0003-4702-7561,sname='Shivaei']{Irene Shivaei}
\affiliation{Centro de Astrobiología (CAB), CSIC-INTA, Carretera de Ajalvir km 4, Torrejón de Ardoz, E-28850, Madrid, Spain}
\email[]{ishivaei@cab.inta-csic.es}

\author[0000-0002-3952-8588,sname='Boogaard']{Leindert A. Boogaard}
\affiliation{Leiden Observatory, Leiden University, PO Box 9513, 2300 RA Leiden, The Netherlands}
\email[]{boogaard@strw.leidenuniv.nl}

\author[0000-0003-0699-6083,sname='D\'{i}az-Santos']{Tanio D\'{i}az-Santos}
\affiliation{Institute of Astrophysics, Foundation for Research and Technology-Hellas (FORTH), Heraklion, GR-70013, Greece}
\affiliation{School of Sciences, European University Cyprus, Diogenes Street, Engomi, 1516 Nicosia, Cyprus}
\email[]{tanio@ia.forth.gr}  

\author[0000-0003-0532-6213,sname='Lofaro']{Cristina M. Lofaro}
\affiliation{Institute of Astrophysics, Foundation for Research and Technology-Hellas (FORTH), Heraklion, GR-70013, Greece}
\email[]{cristinalofaro203@gmail.com}  

\author[0000-0002-7714-688X,sname='Fern\'{a}ndez Aranda']{Rom\'{a}n Fern\'{a}ndez Aranda}
\affiliation{Centro de Astrobiología (CAB), CSIC-INTA, Carretera de Ajalvir km 4, Torrejón de Ardoz, E-28850, Madrid, Spain}
\email[]{rfernandez@cab.inta-csic.es}

\author[0000-0002-6290-3198,sname='Aravena']{Manuel 
Aravena}
\affiliation{Instituto de Estudios Astrof\'{\i}sicos, Facultad de Ingenier\'{\i}a y Ciencias, Universidad Diego Portales, Av. Ej\'ercito 441, Santiago, Chile
}
\affiliation{Millenium Nucleus for Galaxies (MINGAL)}
\email[]{}

\author[0000-0002-2662-8803,sname='Decarli']{Roberto Decarli}
\affiliation{INAF Osservatorio di Astrofisica e Scienza dello Spazio di Bologna, via Gobetti 93/3, Bologna, 40129, Italy}
\email[]{roberto.decarli@inaf.it}  

\author[0000-0003-4268-0393,sname=‘Inami’]{Hanae Inami}
\affiliation{Hiroshima Astrophysical Science Center, Hiroshima University, 1-3-1 Kagamiyama, Higashi-Hiroshima, Hiroshima 739-8526, Japan}
\email[]{hanae@hiroshima-u.ac.jp}

\author[0000-0003-4528-5639]{Pablo G. P\'erez-Gonz\'alez}
\affiliation{Centro de Astrobiología (CAB), CSIC-INTA, Carretera de Ajalvir km 4, Torrejón de Ardoz, E-28850, Madrid, Spain}
\email{pgperez@cab.inta.csic.es}

\author[0000-0003-1151-4659,sname='Popping']{Gerg\"{o} Popping}
\affiliation{European Southern Observatory, Karl-Schwarzschild-Str. 2 D-85748 Garching bei München, Germany}
\email[]{gpopping@eso.org}  

\author[orcid=0000-0001-5434-5942,sname='Van der Werf']{Paul P.~van der Werf} 
\affiliation{Leiden Observatory, Leiden University, PO Box 9513, 2300 RA Leiden, The Netherlands}
\email{pvdwerf@strw.leidenuniv.nl}







\begin{abstract}

We present spatially resolved spectroscopy with JWST/MIRI MRS of a representative sample of normal star-forming galaxies at $z\sim1.1$ as part of the PAHSPECS program. To extract emission from Polycyclic Aromatic Hydrocarbon (PAH) features, we forward model the data cubes with non-parametric spatial distributions, accounting for convolution with the PSF. With this method we are able to recover accurate spatial profiles of the 3.3 $\mu$m, 6.2 $\mu$m, 7.7 $\mu$m, 11.3 $\mu$m PAHs and [ArII] (6.98 $\mu$m) emission and produce PAH ratio maps at cosmic noon. From the PAH ratio maps we find that PAHs become larger and more neutral with increasing galactocentric radius, which is the opposite of trends in local galaxies, indicating radial ISM gradients in normal star-forming galaxies are different at cosmic noon. Through spatially resolved SED fitting of HST and JWST photometry we measure the UV radiation field hardness through the intrinsic ratio of UV to optical flux and find the 3.3/11.3 PAH ratio to decrease with increasing hardness and the 11.3/7.7 to increase. This may suggest photo-destruction of small/ionized PAHs is driving the observed PAH ratio trends and may explain the overall lower 3.3/11.3 PAH at cosmic noon compared to the local universe. This work demonstrates that PAH properties hold crucial information on the resolved ISM physics of galaxies at cosmic noon.

\end{abstract}

\keywords{\uat{Galaxies}{573} \uat{Interstellar medium}{847}}


\section{Introduction}


At the ``cosmic noon'' ($z\sim1-3$), a majority of star-formation is obscured by dust \citep[e.g.][]{Riguccini2011, Zavala2021, Algera2023}. At this critical epoch of galaxy evolution, the properties of the interstellar medium (ISM) are known to be different than the local universe \citep[see][for a review]{Schreiber2020}, with lower metallicities and larger molecular gas reservoirs than local galaxies \citep[e.g.][]{Tacconi2013, Tacconi2018}. Rest frame mid-infrared observations are key to understanding ISM properties, however observing at such wavelengths beyond the local universe is significantly challenging.

A particularly strong observational test of dust properties comes from Polycyclic Aromatic Hydrocarbons (PAHs), which emit strongly in the infrared.
PAHs are the smallest dust grains in the ISM bridging the gap between the molecular and dust phase. These molecules consist of numerous aromatic rings of 20-1000 carbon atoms bonded with hydrogen, where, upon absorption of UV photons from predominantly young stars, will bend and stretch causing broad emission features in the infrared \citep[3-17 $\mu$m; e.g.][]{Tielens2008, Li2020}. Due to their excitation by young stars, PAH emission has been shown to correlate with the star-formation rate (SFR) of galaxies \citep[e.g.][]{Rigopoulou1999, Peeters2004, Calzetti2007, Shipley2016}, as well as both atomic and molecular gas \citep[e.g.][]{Regan2006, Cortzen2019, Leroy2023, Shivaei2024b, Chown2025}.

The strength of PAH emission relative to the dust continuum can be used to measure the fractional contribution of PAHs to total dust while the ratios of different PAH bands provides constraints on the fraction of small vs large and ionized vs neutral contributions to the PAH population \citep[e.g.][]{Draine2007, Draine21, Rigopoulou2021}. Small and ionized PAHs are thought to be more susceptible to destruction through photo-destruction \citep[e.g.][]{Leach1986, Voit1992, Lebouteiller2007, O'Dowd2009} or shocks \citep[][]{Micelotta2010, Zhang2022}. Additionally PAHs are thought to form in either ``top-down'', where large PAHs form first, or ``bottom-up'' processes, where small PAHs form first. PAHs may also coagulate to form bigger molecules/grains in the ISM, though if they collide with too large relative speeds the interaction may instead result in shattering. Shattering may also form PAHs from larger carbonaceous dust grains \citep[e.g.][]{Jones1996, Seok2014}. Therefore PAH ratios can provide constraints on dust formation and destruction mechanisms.

The PAH fraction is known to be dependant on gas-phase metallicity \citep[e.g.][]{Engelbracht2008, Hunt2010, Sandstrom2012, Shivaei2017, Shivaei2024, Whitcomb2024, Whitcomb2025, Lai2025}, with a sharp drop in the PAH fraction below $\sim 30\%$ solar. Additionally PAHs are found to be smaller and more neutral at lower metallicity which is thought to be due to inhibited growth, due to lack of available carbon \citep[e.g.][]{Hunt2010, Sandstrom2012, Whitcomb2024}, resulting in a smaller PAH distribution.


On spatially resolved scales, local galaxies typically show a negative gradient in metallicity \citep[e.g.][]{Vila-Costas1992, Moustakas2010, Sanchez2014}, consistent with a shift towards smaller PAHs at large radii \citep[][]{Whitcomb2024, Whitcomb2025, Baron2025, Koziol2026, Lofaro2026a, Lofaro2026b}, however galaxies at higher redshifts show a greater diversity of metallicity gradients, often with flat gradients \citep[e.g.][]{Curti2020}. In addition to different formation mechanisms, feedback processes, such as from an Active Galactic Nucleus (AGN), can alter the PAH population. In particular the presence of an AGN in local galaxies, results in more neutral PAHs \citep[e.g.][\textcolor{xlinkcolor}{}]{Zhang2024, Garcia-Bernete2022c, Garcia-Bernete2024b, Rigopoulou2024, Lofaro2026a, Lofaro2026b, Donnan2026}, particularly in regions where AGN outflows impact star-forming regions, potentially explained by destruction of the ionized PAHs in AGN outflows.

While there has been much work at cosmic noon relying on photometry \citep[e.g.][]{Reddy2010, Shipley2016, Shivaei2017, Shivaei2024, Shivaei2024b}, revealing the fraction of dust that PAHs constitute, spectroscopy is required to properly measure PAH band ratios and therefore reveal key properties of the dust at this epoch. In particular, photometry can suffer from poor continuum subtraction, overlapping PAH features and even line contamination. Recent MIRI LRS (Low Resolution Spectroscopy) of massive infrared luminous galaxies at cosmic noon \citep[$z\sim0.7-2.5$][]{McKinney2025} have revealed lower 3.3 $\mu$m PAH fluxes relative to the 11.3 $\mu$m, suggesting larger grains compared to local galaxies at similar infrared luminosities. However LRS is a slit that only covers up to $\sim 14\mu$m ($\lesssim7\mu$m at $z\gtrsim1$), relying on Spitzer to measure the 11.3 $\mu$m PAH. With MIRI MRS (Medium Resolution Spectroscopy) there is a higher wavelength coverage up to  $\sim 28\mu$m allowing one to measure the 11.3 $\mu$m PAH at $z\sim1.1$. Moreover MIRI MRS is an integral field unit (IFU), which enables spatially resolved studies of the PAH properties, which was not possible to do at $z\gtrsim0.5$ before the advent of JWST. 

In this work we present JWST MIRI MRS observations of five main sequence galaxies at $z\sim1.1$ as part of the PAHSPECS program (GO 5279, P.I. I. Shivaei $\&$ L. Boogaard $\&$ T. D\'iaz Santos). These galaxies were selected from the ALMA Spectroscopic Survey in the Hubble Ultra Deep Field (ASPECS) sample \citep{Walter2016, Aravena2019, Aravena2020, Boogaard2019, Boogaard2020, Decarli2019, Decarli2020, Gonzalez2019, Gonzalez2020}, which performed spectral scans of the 1.2mm and 3mm ALMA bands to detect CO emission and dust continuum emission of galaxies across cosmic time up to $z\sim4$ in the Hubble Ultra Deep Field (HUDF), providing a flux-complete sample. The PAHSPECS sample are all the galaxies at $z\sim1.1$ within the parent ASPECS sample, providing a representative sample of typical star-forming galaxies at this epoch.

In Section \ref{sec:Obs} we present the observations and data reduction of the five galaxies. In Section \ref{sec:Methods} we describe how we model the data cubes, accounting for PSF convolution to accurately infer PAH maps. In Section \ref{sec:Results} we present maps of the PAH emission and their ratios. Finally in Section \ref{sec:Discuss} we discuss the results in the context of the literature and the implications for the evolution of dust since cosmic noon. Throughout this work, where needed, we assume $\Lambda$CDM cosmology with $H_0 = 70$ km s$^{-1}$ Mpc$^{-1}$, $\Omega_m = 0.27$, $\Omega_{\Lambda} = 0.73$.

\section{Observations and Data Reduction}
\label{sec:Obs}
The five galaxies in our sample shown in Table \ref{tab:Sample}, are selected as all the galaxies at $z\sim1.1$ from the flux-limited ASPECS sample, which consists of main-sequence galaxies. A redshift of  $z\sim1.1$ is the highest possible redshift at which the 11.3 $\mu$m PAH is visible within the wavelength range of MIRI MRS. All of these galaxies show CO detections apart from ASPECS-C20, which is a galaxy with a low SFR. Only ASPECS-11 shows no 1mm dust continuum while APSECS-C20 shows no 3mm dust continuum detection. Both the 1mm and 3mm ID's are givien in table \ref{tab:Sample}. A full overview of the sample, including broadband SED modeling can be found in \citet{Shivaei2026}.

\begin{table*}
\centering
  \caption{PAHSPECS Sample}
  \label{tab:Sample}
    \def\arraystretch{1.2}
    \setlength{\tabcolsep}{4pt}

    \begin{threeparttable}
  \begin{tabular}{cccccc}
  
    \hline

    Name & RA & Dec & $z$ & 1mm ID &  3mm ID \\
    (1) & (2) & (3) & (4) & (5) &  (6) \\

    \hline
    ASPECS-6 & 03:32:39.87 & -27:47:15.2 & 1.0952 & 1mm.C16 & 3mm.06  \\
    ASPECS-11 & 03:32:39.81 & -27:46:53.5 & 1.0964 & - & 3mm.11 \\
    ASPECS-14 & 03:32:34.85& -27:46:40.6 & 1.0982 & 1mm.C25 & 3mm.14  \\
    ASPECS-15 & 03:32:36.48& -27:46:31.8 & 1.0931 & 1mm.C12 & 3mm.15 \\
    ASPECS-C20 & 03:32:35.77 & -27:46:27.6 & 1.0963 & 1mm.C20 & -  \\

    \hline
  
  \end{tabular}
\begin{tablenotes}
    \item[] Column (1): Source name. Column (2): Right Ascension. Column (3): Declination. Column (4): Redshift. Column (5): ALMA ASPECS 1 mm continuum ID \citep{Aravena2020, Gonzalez2020}. Column (6): ALMA ASPECS 3 mm CO ID \citep[][]{Boogaard2019, Gonzalez2019}.
    \end{tablenotes}
  \end{threeparttable}
 \end{table*}




The MRS data for each target was a single pointing with 4 dithers and dedicated background observations. The MIRI MRS data were reduced using calibration version 1.17.0 with CDRS version 12.0.5. We use the standard JWST reduction pipeline with some modifications. In particular, during stage 2 of the pipeline we include the residual fringe correction as well as additional cosmic ray mitigation to remove additional artifacts in the data caused by cosmic showers. The background exposures were taken with the number of groups and integrations matching the longest exposure observations but with a 2-point dither, matching the recommended observing strategy. We initially tried subtracting a single background value, averaging over the background frames to improve signal to noise however we found that doing the background subtraction for each spaxel yielded better results.
For more details about the data reduction steps see \citet{Shivaei2026}.

Cubes are aligned and rotated into the standard sky alignment (north is up and east is to the left) before any analysis.
We additionally estimate errors for each spaxel after creating sub-cubes for each feature of interest. We do this by directly measuring the error from the scatter of the data.  For the PAH features we first use a median filter (width of 25 wavelength pixels) to smooth the spectrum of each spaxel. The standard deviation is then measured after subtracting the smoothed spectrum to calculate an error map. For emission lines we mask the line instead of smoothing and simply measure the standard deviation of the points on either side of the emission line. We find that this data driven calculation of the errors returns $\sim10\%$ larger errors than the pipeline, potentially capturing additional sources of uncertainty. The majority of the noise dominating the uncertainties is the telescope background, however for the 11.3 $\mu$m PAH, which is observed in channel 4, the degradation of the detector at long wavelengths will also contribute to the noise.

As the key aim of this work is to create PAH ratio maps, it is important that each data cube has proper astrometric alignment. After consulting the JWST help desk, the WCS produced by the pipeline was determined to be well aligned, with the only possible offsets being due to the filter wheel positioning introducing offsets up to $\sim 30$ milliarcseconds, much less than a pixel element \citep[][]{Patapis2024}.



\section{Extracting Emission Features}
\label{sec:Methods}



To generate maps of specific spectral features, the continuum due to thermal dust emission and/or stellar continuum needs to be subtracted. For PAH features, their broad shape means subtracting a continuum is not trivial, and therefore typically requires modeling the full mid-IR spectrum \citep[e.g.][]{Smith2007, Donnan24a, Diaz-Santos2025, VanDePutte2025}. In \citet{Lofaro2026c}, this approach is taken to model the integrated spectra of the PAHSPECS galaxies. However to produce spatially resolved maps, extracting spectra of each spaxel is not feasible due to the low signal-to-noise of each spaxel. We instead opt to forward model the data cube, to determine the spatial distribution of each PAH feature, including PSF convolution. By taking this approach, we use the spatial dimensions and therefore maximize the information held within the data to extract high quality PAH feature maps. 

While these galaxies are spatially resolved, they are at very similar scales to the PSF. This means obtaining accurate spatial distributions as well as ratios of different emission features requires careful modeling, accounting for the PSF, and how it changes with wavelength. 


In this work we extract the spatial distribution  maps of emission features by forward modeling the cube of a given feature and its continuum. Not only does this account for the different PSF sizes for different emission features but this method maximizes the information within the data by fitting the cube rather than each spaxel independently. Given the relatively low signal to noise of our high-$z$ sample compared to local galaxies \citep[e.g.][]{Rigopoulou2024}, particularly for the 11.3 $\mu$m PAH, which for our targets is observed in channel 4 at $\sim24\mu$m, fitting each spaxel independently to subtract a continuum is not feasible. 

In this work, we model the 3.3 $\mu$m PAH, 6.2 $\mu$m PAH, the 7.7 $\mu$m PAH, the 11.3 $\mu$m PAH and [ArII] (6.985 $\mu$m, ionization potential 27.6 eV). While it is theoretically possible to forward model the entire cube from 5-28 $\mu$m (observed wavelength), doing so is highly computationally expensive. We instead first create sub-cubes of each feature from the data which has $\mu$ and $\nu$ spatial pixels and $m$ spectral pixels. 
As each of the 4 MIRI MRS channels have different pixel scales, the model is generated on a pixel grid matching the data pixel scale of channel 3 ($0.2''\times0.2''$) which contains the 6.2 $\mu$m and 7.7 $\mu$m PAHs. We choose to generate the model PAH emission maps on the same pixel grid for all the features, namely $0.2''\times0.2''$. This is the scale of the data cube containing the most emission features analysed in this work, namely the 6.2 $\mu$m PAH, 7.7 $\mu$m PAH and the [ArII] line and therefore minimizes the number of features that need to be interpolated onto the data. As the data cubes of the 3.3 $\mu$m and 11.3 $\mu$m PAHs are on different pixel scales, appearing in channel 1 and channel 4 respectively, the model PAH maps are generated on a different pixel scale to the data. Therefore, during the fitting process, after convolution with the PSF, they need to be interpolated onto the pixel scale of the data to calculate the log likelihood. 

The advantage of modeling all the features on the same pixel scales is that maps of the ratios can be made without further interpolation, producing cleaner ratio maps.

The model consists of a feature template, $T(\lambda)$, plus a continuum as a function of wavelength, both of which have some spatial distribution to produce a model cube before convolving with a PSF cube to fit the data. Before convolution the model cube, $M\left(x, y, \lambda \right)$, has the form
\begin{equation}
\label{eqn:Model}
    M\left(x, y, \lambda \right) = F(x,y)T(\lambda) + C_0(x,y)[\lambda - \lambda_0] + C_1(x,y), 
\end{equation}
where the feature template, $T(\lambda)$, is normalised and scaled by $F(x,y)$, which is the spatial distribution of the emission feature. 
The continuum has the form $C_0(x,y)[\lambda - \lambda_0] + C_1(x,y)$, a linear function of wavelength with a mean value of $C_1(x,y)$ which is the spatial distribution of the mean continuum flux, where $\lambda_0$ is the midpoint of the sub-cubes, while $C_0(x,y)$ describes the slopes of the continua of each spaxel. A linear function is a reasonable approximation for the continuum over a short wavelength range, especially considering the low signal to noise of the continuum in this data.

We also include a velocity parameter for each spaxel, $V(x,y)$, to account for velocity shifts due to rotation within the galaxy. While it is possible to measure kinematics of PAH features in local galaxies \citep[][]{Donnan24b, Donnan2026}, we find that the signal to noise is too low to extract kinematics for the higher redshift PAHSPECS targets. For this reason we do not include a line spread function (LSF).

\begin{figure}
        \includegraphics[width=\columnwidth]{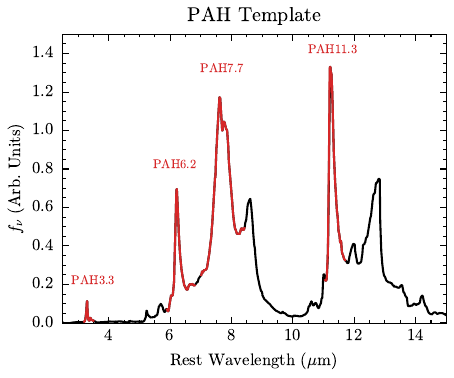}
    \caption{PAH template generated from the average spectrum of 12 star-forming regions in the local LIRGs, NGC 3256 and NGC 7469 \citep[][]{Donnan24a, Rigopoulou2024}. The spectrum has been continuum subtracted by fitting with \textsc{SPIRIT} \citep{Donnan24a} and smoothed. The bands highlighted in red show the individual templates used in this work to fit to the PAHSPECS data.}
    \label{fig:PAHTemplate} 
\end{figure}

To model the PAH features we use a fixed template as any variations in the feature shape are too small to be detected at the signal to noise of our data. Moreover, by using a template, we include the broad wings of the PAH features and the pseudo-continuum due to the wings of the features blending, which is particularly strong for the 7.7 $\mu$m PAH. At the signal to noise of the data, fitting the wings/pesudo-continuum would be extremely challenging. To generate the template, we take the average of 12 continuum subtracted spectra of star-forming regions in two local LIRGs, NGC 3256 and NGC 7469, described in \citet{Donnan24a}, which show typical PAH properties of star-forming regions \citep[][]{Rigopoulou2024}. We then smoothed the average continuum subtracted spectrum with a median filter to remove any emission lines. We find these templates to be good fits for the shape of the PAH bands of the ASPECS galaxies. The templates,  $T(\lambda)$, for the 3.3 $\mu$m, the 6.2 $\mu$m, 7.7 $\mu$m and 11.3 $\mu$m PAH bands are shown in Fig. \ref{fig:PAHTemplate}, highlighted in red. It is worth noting that the 3.3 $\mu$m template also includes the aliphatic emission complex at $\gtrsim$3.4$\mu$m. 

To ensure that the PAH fluxes inferred by using templates are comparable to fitting Drude profiles to a full integrated spectrum using methods such as \textsc{CAFE} \citep[][]{Diaz-Santos2025} or \textsc{SPIRIT} \citep[][]{Donnan24a}, we calculate the factor difference between the integrated flux of the template (in Fig. \ref{fig:PAHTemplate}) and the integrated flux of a fit to the full template using Drude profiles. We then correct the PAH maps by this factor. For example, in the case of the 3.3 $\mu$m PAH feature, this reduces the flux, as the 3.4 $\mu$m aliphatic feature is also included in the template. The factors are 0.84, 0.71, 1.02, 1.17 for the 3.3, 6.2, 7.7, and 11.3 $\mu$m PAHs respectively.

The model cube, $M\left(x, y, \lambda \right)$, is then convolved with a PSF cube which is then fitted to the data. We use WebbPSF\footnote{\url{https://stpsf.readthedocs.io/en/latest/intro.html}} to generate a PSF for each channel of the sub-cubes. It is worth noting that while WebbPSF simulates PSFs from the mirror geometry, wavefront error maps and detector pixel sampling, it does not simulate the effects of the image slicer. Considering we have no strong point sources in our data that would need extremely precise PSFs, the WebbPSF outputs are suitable for this work.

The PSF cubes are all generated on the same pixel scale, equal to that of the data in channel 3. During the fitting process, each frame of the model cube, given by equation \ref{eqn:Model}, is convolved with each frame of the PSF cube at each iteration during the fitting procedure. If the feature being modelled is the 3.3 $\mu$m PAH or the 11.3 $\mu$m PAH, the model cube after convolution is also interpolated onto the pixel scale of the data in order to calculate the likelihood, $\ln \textrm{Prob}$ (equation \ref{eqn:prob}), at the current iteration. 


The only clear emission line in the data is [ArII] (6.985 $\mu$m), which is present in all the targets. Rather than using a template for this feature, we model the profile with a simple Gaussian, where we include an additional parameter to control the width of the Gaussian. Rather than allowing each spaxel to have a different velocity dispersion, for simplicity (and due to the relatively low signal to noise of the data), we use the same velocity dispersion ($\sim 100$ km s$^{-1}$) and thus the width of the Gaussian for each spaxel. 

The spatial distribution of the feature and underlying continuum are non-parametric in this model where 
the value of the PAH flux at each spaxel, $F(x,y)$, the continuum flux and slope, $C_1(x,y)$, $ C_0(x,y)$, and the velocity of each spaxel, $V(x,y)$, are all free parameters. For a pixel grid of $\mu$ and $\nu$ spatial pixels, each of the $\mu\times\nu$ spaxels have four parameters, for a total of $4\times\mu\times\nu$ parameters. Because of the structure of galaxy disks, we don't expect that the resulting maps will have random spatial distributions. To improve the S/N, we make use of that fact by including a regularization term where each spaxel cannot vary too much from each neighbouring spaxel. The regularisation term is included in the log-likelihood during the fitting procedure. This acts as a penalty where smooth spatial distributions incur a lower penalty. The best fit solution is therefore a compromise between the maximum likelihood and lowest penalty. We follow \citet{Donnan24a}, inspired by the regularisation on the non-parametric star formation histories in \textsc{pPXF} \citep[][]{Cappellari2017, Cappellari2022} and use the square of the Laplacian as a measure of the smoothness of the spatial distribution of the emission feature given by

\begin{multline}
\label{eqn:Laplace}
    P_{F} = \sum_{x} \left(F_{x-1, y}  - 2 F_{x, y} + F_{x+1, y}\right)^2 \\
    + \sum_{y} \left(F_{x, y-1}  - 2 F_{x, y} + F_{x, y+1}\right)^2,
\end{multline}
where $F(x,y)$ is the spatial distribution of a given emission feature. The total probability, $\ln \textrm{Prob}$, maximized during the fit has the form
\begin{equation}
\label{eqn:prob}
  \ln \textrm{Prob} = -\chi^2 - \Gamma\left( P_F + P_V + P_C \right) + \textrm{const.},
\end{equation}
where $P_F$, $P_V$, and $P_C$ are the penalty factors for the feature flux map, velocity map and continuum flux map respectively. The constant term is a normalisation term such that the left hand side of the equation is a probability. The strength of the regularisation is controlled by the constant $\Gamma$. If the value is too low, the spatial distributions of the emission features and continuum will be noisier while if the value of $\Gamma$ is too large, it will be too smoothed, leading to overestimating the size of the emitting region. To quantify this effect and choose the most appropriate value for $\Gamma$, we conducted testing in Appendix \ref{sec:Reg}. We calculate the half-light radius for the inferred spatial distribution of the 6.2 $\mu$m PAH of ASPECS-6 for different values of $\Gamma$.
We find that different features require different values of $\Gamma$, where the optimum values are $\Gamma=1$ for the 6.2 $\mu$m and 7.7 $\mu$m PAH features, $\Gamma=0.01$ for the 11.3 $\mu$m PAH, $\Gamma = 10$ for the 3.3 $\mu$m PAH and $\Gamma = 100$ for [ArII] (6.98 $\mu$m). The higher regularisation for the shorter wavelength features likely reflects the change in spatial resolution with wavelength where the longer wavelength 11.3 $\mu$m PAH has the strongest weight per arcsecond while the 3.3 $\mu$m PAH has the weakest. Moreover, despite the [ArII] line having the same spatial resolution as the 6.2 $\mu$m and 7.7 $\mu$m PAHs, the feature is significantly narrower and thus has a lower weight on the total log-likelihood than a broad feature.



We achieve a best fit solution by maximising equation (\ref{eqn:prob}). To obtain uncertainties on the inferred flux/velocity maps, we use a Monte Carlo method, where we resample the flux data assuming a normally distributed errors on the flux measurements and re-fit the model. By repeating this process 100 times we build up 100 samples of the best fit solution where we take the mean and standard deviation to give the value and error of each parameter and thus the flux/velocity maps.

\section{Results}
\label{sec:Results}
\subsection{PAH and Continuum Maps}
\begin{figure*}
\centering
        \includegraphics[width=16cm]{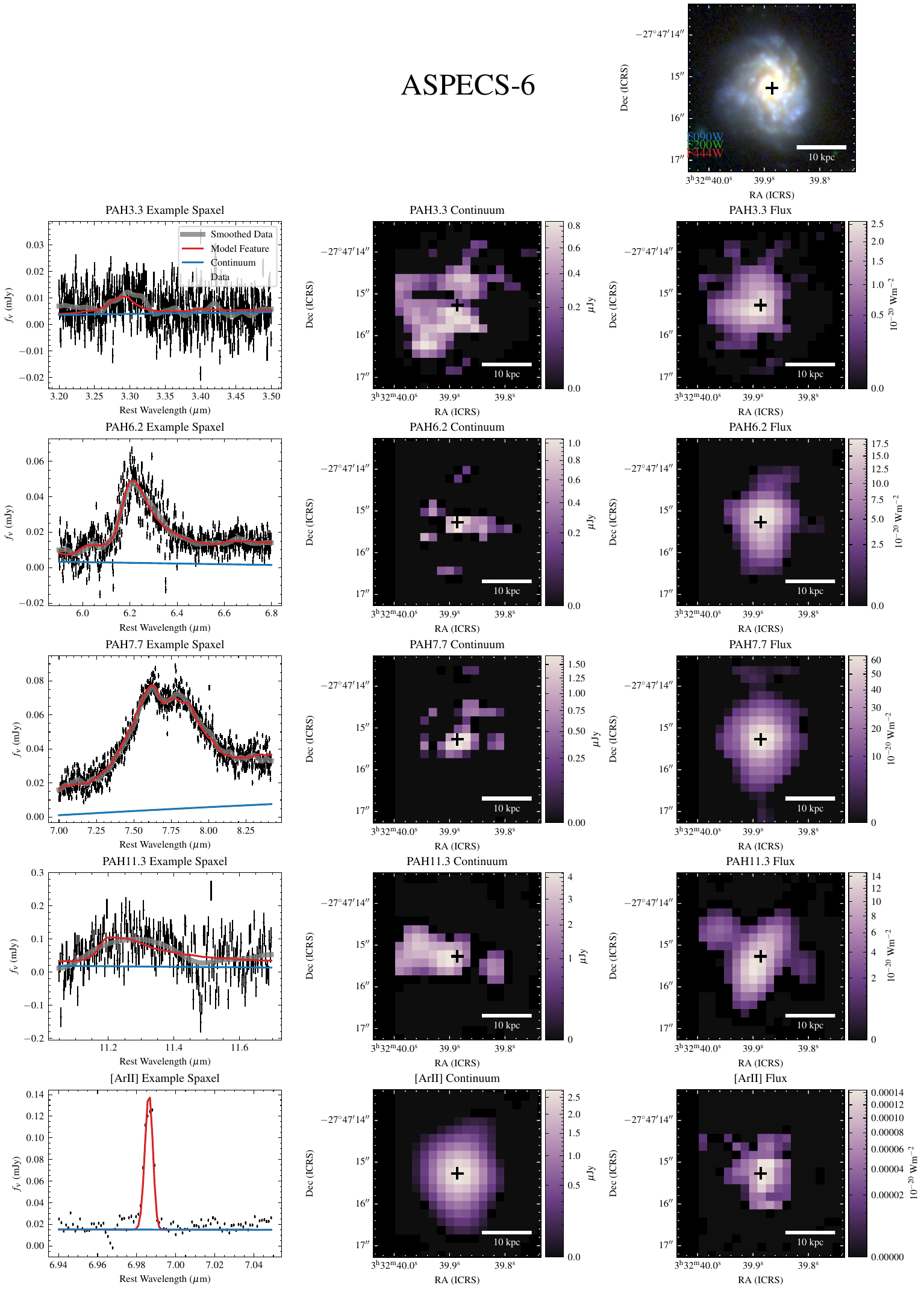}
    \caption{Results from cube fitting ASPECS-6 where each row shows the different emission features and the continuum at that wavelength. We show a NIRCAM color image in the top right panel. The first column shows the brightest spaxel with the best fit model feature and continuum. The grey line shows the data smoothed with a Gaussian kernel with a width of 15 wavelength elements. The second column shows the model continuum map at the wavelength of the corresponding emission feature. The third column shows the inferred feature map. The black cross marks the center of the galaxy. Note that the continuum for the [ArII] feature is not a true continuum and is instead dominated by the wings of the 6.2 and 7.7 $\mu$m PAH features, hence the discrepant morphology compared to the other continuum maps.
    }
    \label{fig:ASPECS-6} 
\end{figure*}

\begin{figure*}
\centering
        \includegraphics[width=16cm]{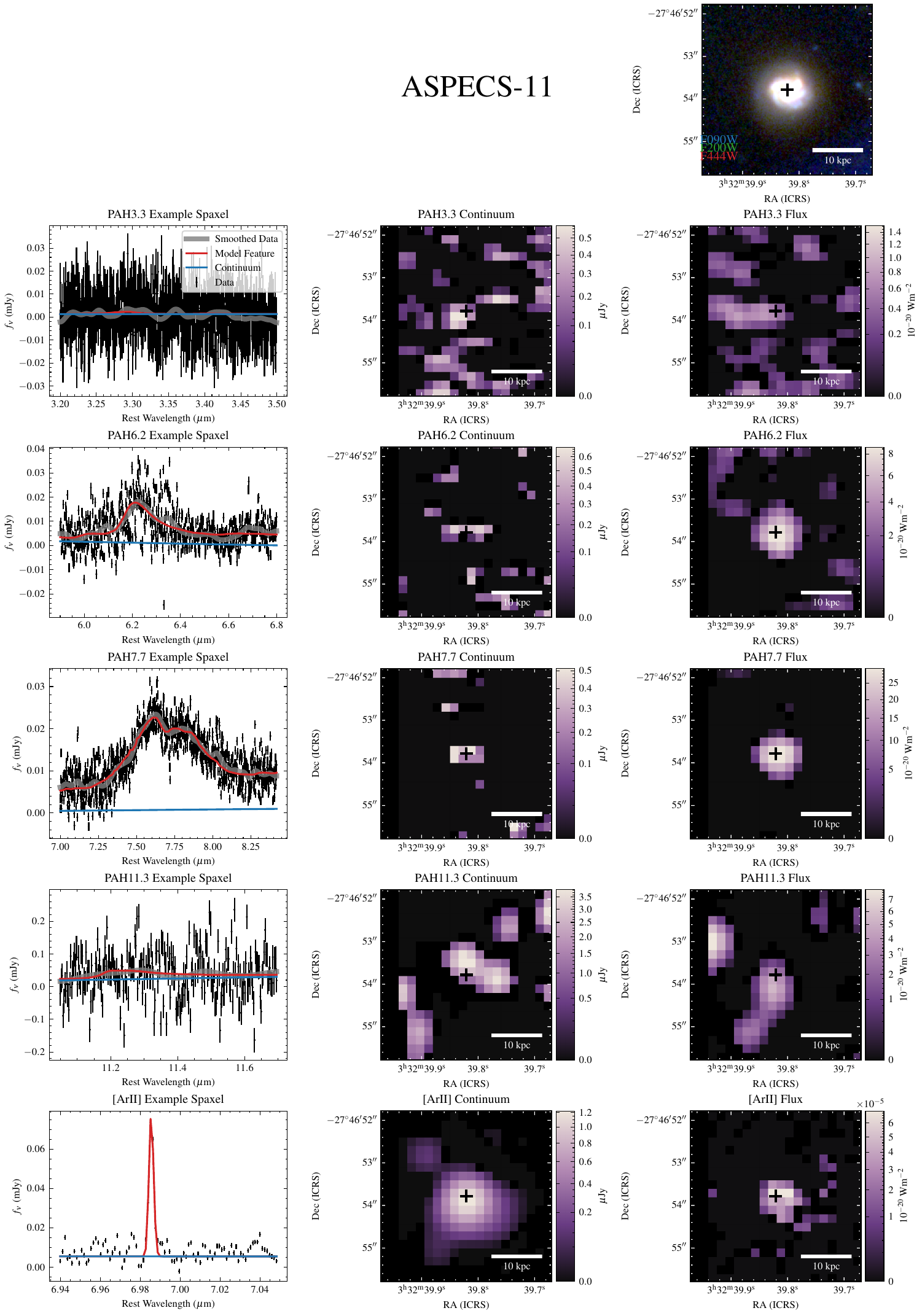}
    \caption{Same as Fig. \ref{fig:ASPECS-6} for ASPECS-11.}
    \label{fig:ASPECS-11} 
\end{figure*}

\begin{figure*}
\centering
        \includegraphics[width=16cm]{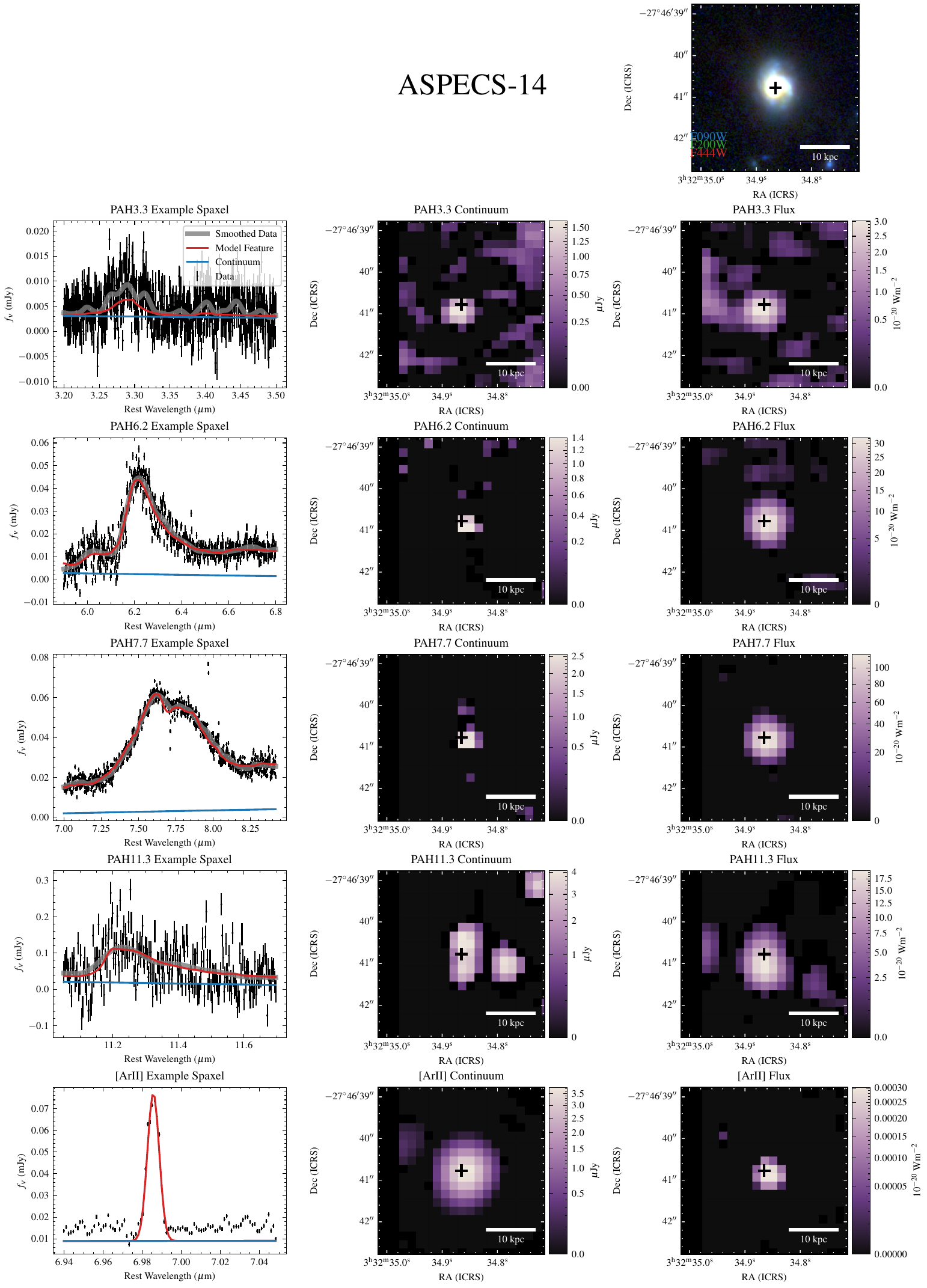}
    \caption{Same as Fig. \ref{fig:ASPECS-6} for ASPECS-14.}
    \label{fig:ASPECS-14} 
\end{figure*}

\begin{figure*}
\centering
        \includegraphics[width=16cm]{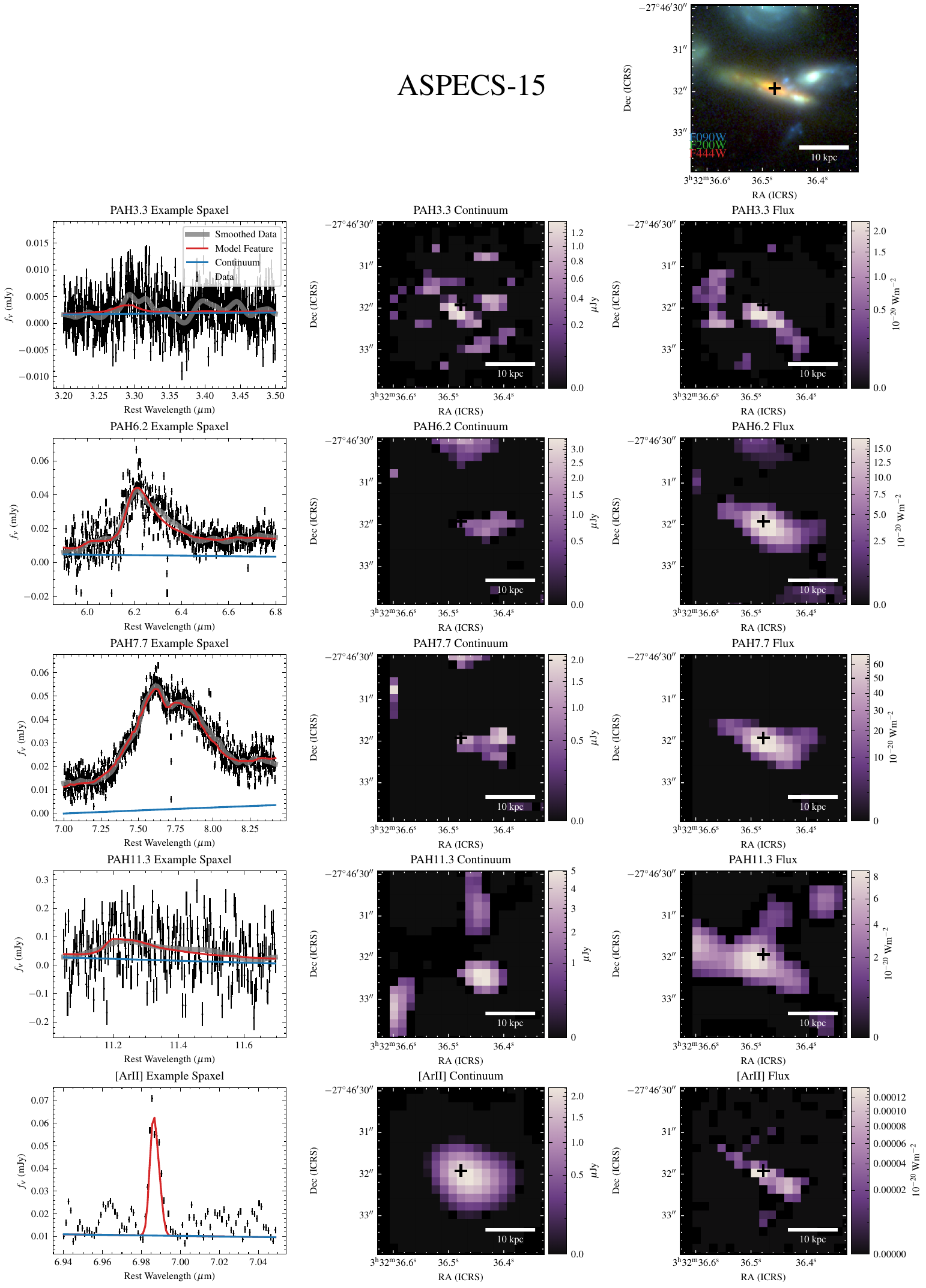}
    \caption{Same as Fig. \ref{fig:ASPECS-6} for ASPECS-15.}
    \label{fig:ASPECS-15} 
\end{figure*}

\begin{figure*}
\centering
        \includegraphics[width=16cm]{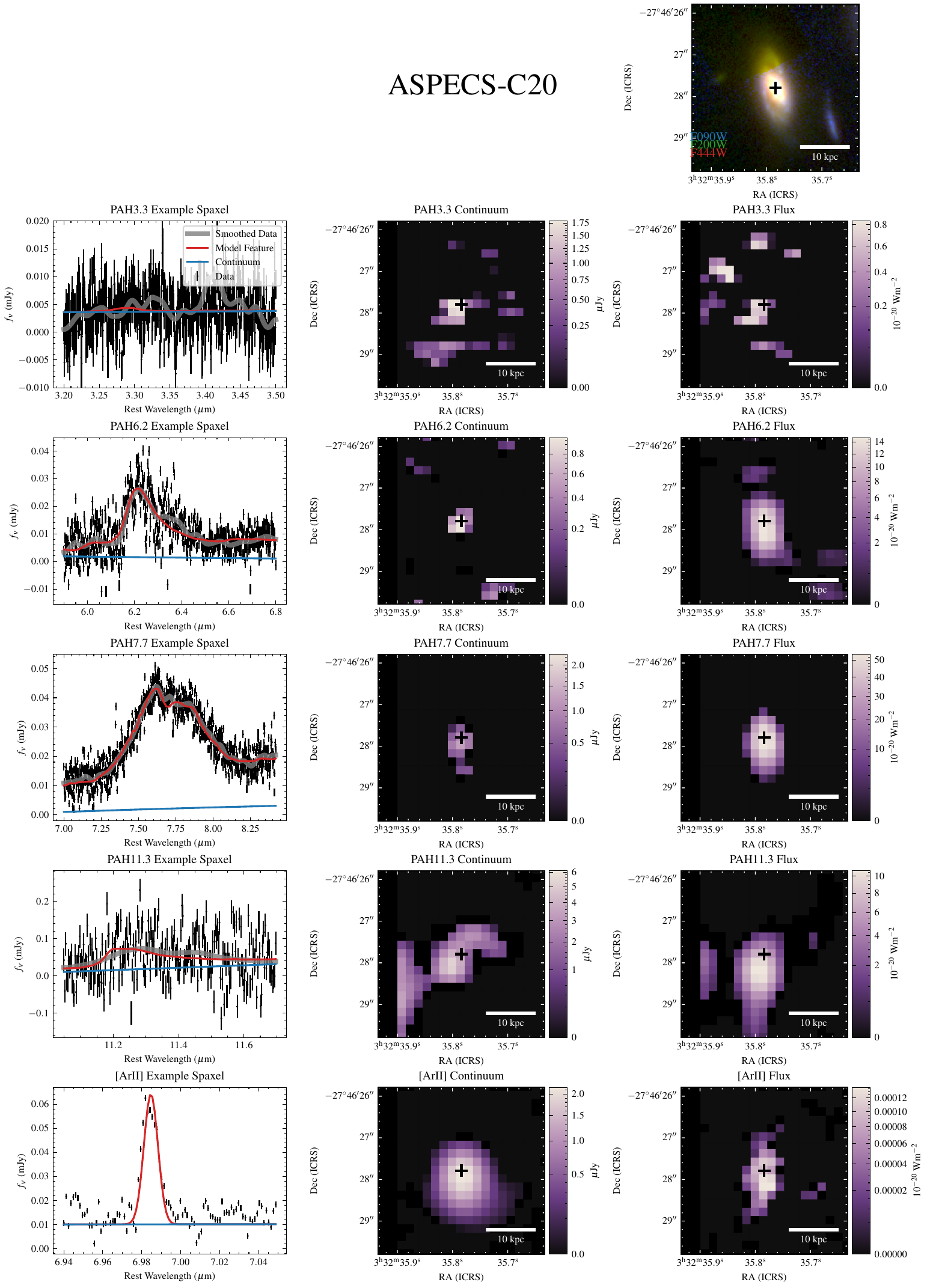}
    \caption{Same as Fig. \ref{fig:ASPECS-6} for ASPECS-C20.}
    \label{fig:ASPECS-C20} 
\end{figure*}

For each of the 5 galaxies we present the results of the fitting of the 3.3, 6.2, 7.7, and 11.3 $\mu$m PAH and the 6.98 $\mu$m [ArII] line in Figures \ref{fig:ASPECS-6}, \ref{fig:ASPECS-11}, \ref{fig:ASPECS-14}, \ref{fig:ASPECS-15}, and \ref{fig:ASPECS-C20}. In these Figures we show the maps of the emission feature (moment 0) and the continuum at the corresponding wavelength, where we only show pixels that have a $>3\sigma$ detection. 
The leftmost column shows an example spaxel with the best fit model at that spaxel. These panels show the strength of fitting the cube rather than individual spaxels as they can be very noisy, for example the 3.3 $\mu$m and 11.3$\mu$m PAH features, making the measured flux unreliable and the continuum completely unconstrained. By fitting in 3D, we maximize the spatial information and so we are able to constrain the PAH and continuum maps as demonstrated in the top panels of Fig. \ref{fig:ASPECS-6}, even when the individual spaxels are noisy.

We find that all the emission features show clear detections for three galaxies in our sample, namely ASPECS-6, ASPECS-14, ASPECS-15. For ASPECS-11, we find the 3.3 $\mu$m PAH shows no robust detection, even in the integrated spectrum \citep[][]{Lofaro2026c}. Similarly, in ASPECS-C20 the 3.3 $\mu$m PAH is very weak, with a marginal detection at the position of the galaxy.

The continuum at wavelengths $\gtrsim4\,\mu$m, arising from thermal emission of hot dust grains, is much fainter than the PAH features for all the galaxies and the morphology is different. In particular, the hot dust continuum is more compact than the PAH emission. This is partly due to the signal to noise being low and so we lose any extended dust continuum emission however a compact dust continuum is consistent with local galaxies \citep[e.g.][]{Diaz-Santos2010, Diaz-Santos2011, Pathak2024}, likely reflecting a higher dust temperature in the nucleus and thus a higher energy density of photons. This makes sense considering the higher star-formation rate in the nucleus (see Fig. \ref{fig:Pipes}), and the higher number density of stars, leading to hotter dust in the nucleus. We confirm this via MIRI imaging, where the F1000W filter traces the $\sim5\mu$m continuum which shows more compact emission than the F1500W filter or F770W filter, after convolving to the same PSF scale, both of which are dominated by PAH emission at $z\sim1.1$. For ASPECS-6, the continuum at 6.2 $\mu$m and 7.7 $\mu$m appears compact but potentially elongated east to west and therefore may trace a bar-like structure in this galaxy. For the most compact sources, namely ASPECS-11 and ASPECS-14, the hot dust continuum is still clearly more compact than the PAH emission.

Unlike the other PAH features, the continuum at 3.3\,$\mu$m contains contribution from starlight, likely more than hot dust, resulting in a different morphology than the hot dust continuum. However we do find a discrepancy here, where for ASPECS-6 for example, the 3.3\,$\mu$m continuum does not peak the centre which is inconsistent with the imaging (as seen in the color image in Fig. \ref{fig:ASPECS-6}). This may suggest the data is simply too noisy at  3.3\,$\mu$m to produce a continuum map. We also check the F560W MIRI image, which traces $\sim2.5\mu$m rest frame emission and it shows a strong nuclear bulge which is missing in our 3.3 $\mu$m continuum image. The slope of this continuum is positive (increases over the wavelength range) which may suggest some hot dust contribution however, considering the signal to noise is very low in this part of the spectrum, we do not consider this continuum map as reliable as the longer wavelength continuum maps.

The [ArII] emission appears relatively compact compared to the PAH emission however it is not as compact as the hot dust continuum. [ArII] has a low ionization potential of 27.6 eV, tracing ionized gas primarily from star-forming regions. In ASPECS-15, the [ArII] emission is notably off-center, peaking towards the south east. This region is coincident with a particularly UV bright region, with blue optical colors.

It is worth noting that the continuum inferred at 7 $\mu$m for [ArII] (6.98 $\mu$m) will contain a significant contribution from the wings of the 6.2 $\mu$m and 7.7 $\mu$m PAHs. This can be seen in the PAH template in  Fig. \ref{fig:PAHTemplate}. The inferred continuum presented in the second column of Fig. \ref{fig:ASPECS-6} therefore is a mix of dust continuum and PAH emission explaining its different spatial structure compared to the continuum inferred for the 6.2 $\mu$m and 7.7 $\mu$m PAHs.


\subsection{PAH Band Ratios}

\begin{figure*}
        \includegraphics[width=\textwidth]{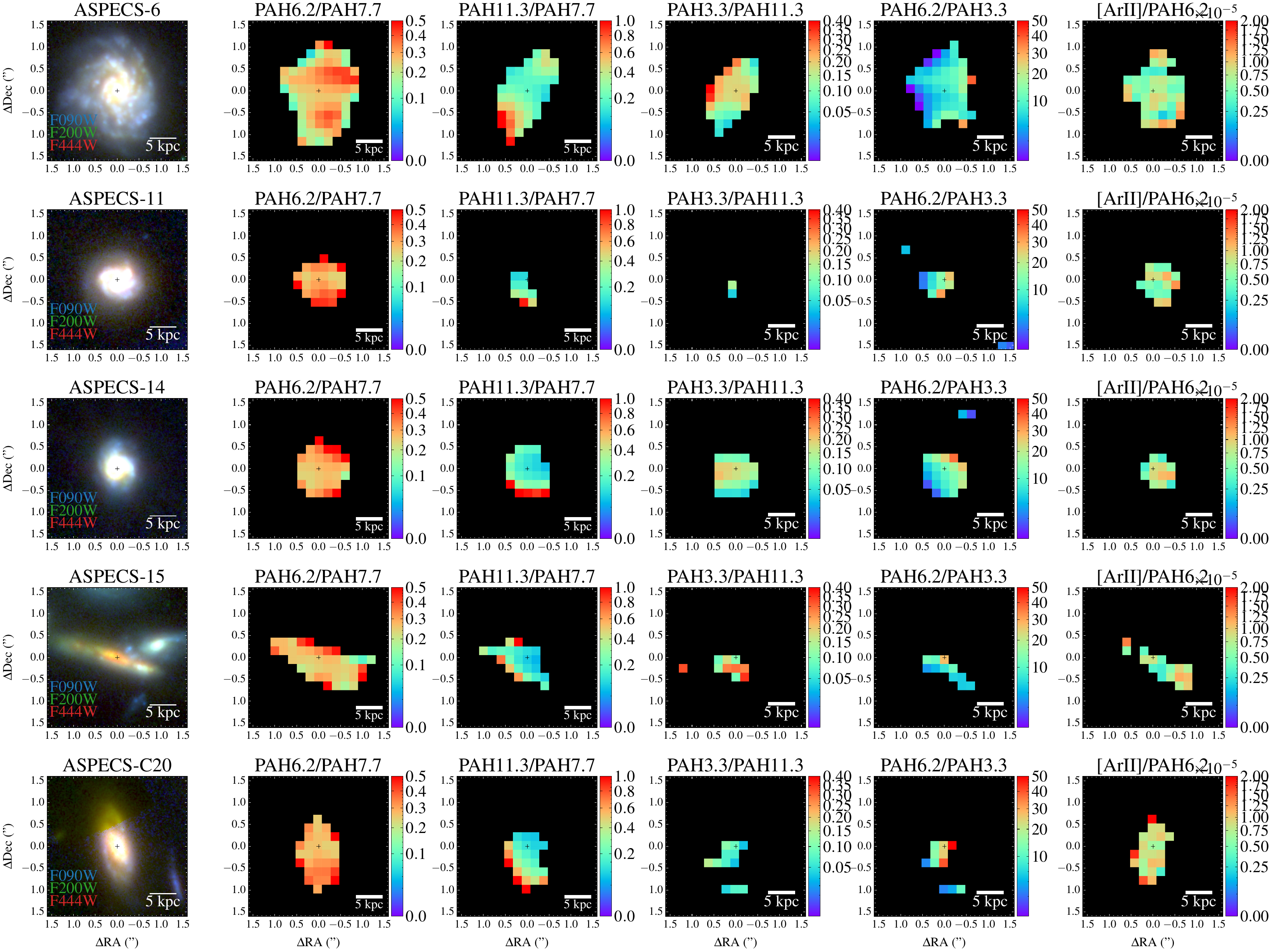}
    \caption{PAH ratio maps for the five PAHSPECS galaxies. We only show pixels that have a $>3\sigma$ measurement of the ratio. The left column shows NIRCAM color images using the F090W, F200W, F444W filters \citep[][]{Rieke2023}. The second column shows the PAH6.2/PAH7.7 ratio. The third column shows the PAH11.3/PAH7.7 ratios. The fourth column shows the PAH3.3/PAH11.3 ratios. The final column shows [ArII]/PAH6.2.}
    \label{fig:PAHRatios} 
\end{figure*}

With the maps inferred for all the targets, we construct maps of different PAH ratios to determine how the PAH properties vary spatially, enabled by accounting for PSF convolution in our modeling. We present maps of the 6.2/7.7, 11.3/7.7, 6.2/3.3 and 3.3/11.3 PAH ratios for all five of the galaxies in Fig. \ref{fig:PAHRatios} alongside color images comprised of NIRCAM filters F090W, F200W, F444W from JADES \citep[][]{Rieke2023}. We find clear spatial variations in the PAH ratios, particularly for ASPECS-6 which is the most extended galaxy in the sample. 

We do not apply any extinction correction to the PAH ratios as these galaxies are not particularly dusty, with maxiumum values of $A_v\sim1$ as presented in section \ref{sec:Ancillary}. In the presence of substantial attenuation, we would expect that the 11.3/7.7 PAH ratio to increase once corrected for extinction, while the 3.3/11.3 and 6.2/7.7 are relatively insensitive to extinction. Extinction of these galaxies is explored further with the integrated spectra in \citet{Lofaro2026c}, where the 11.3/7.7 can change by $\sim10\%$.

It is worth noting that any of the spaxels displayed in the ratio maps are those with robust detections in both PAH features ($>3\sigma$) and so there are other spaxels not included where one feature may be detected but the other was not. 

We caution the reader about the 3.3/11.3 and 6.2/3.3 ratio maps for ASPECS-11 and ASPECS-C20 as the 3.3 $\mu$m PAH does not show a robust detection, particularly when inspecting the 1D spectra, although there are spaxels consistent with the galaxy position in the 3.3 $\mu$m PAH flux map. We include ratios involving the  3.3 $\mu$m PAH feature for ASPECS-11 and ASPECS-C20 for completeness.

\begin{figure*}

    \centering
        \includegraphics[ width=16.5cm]{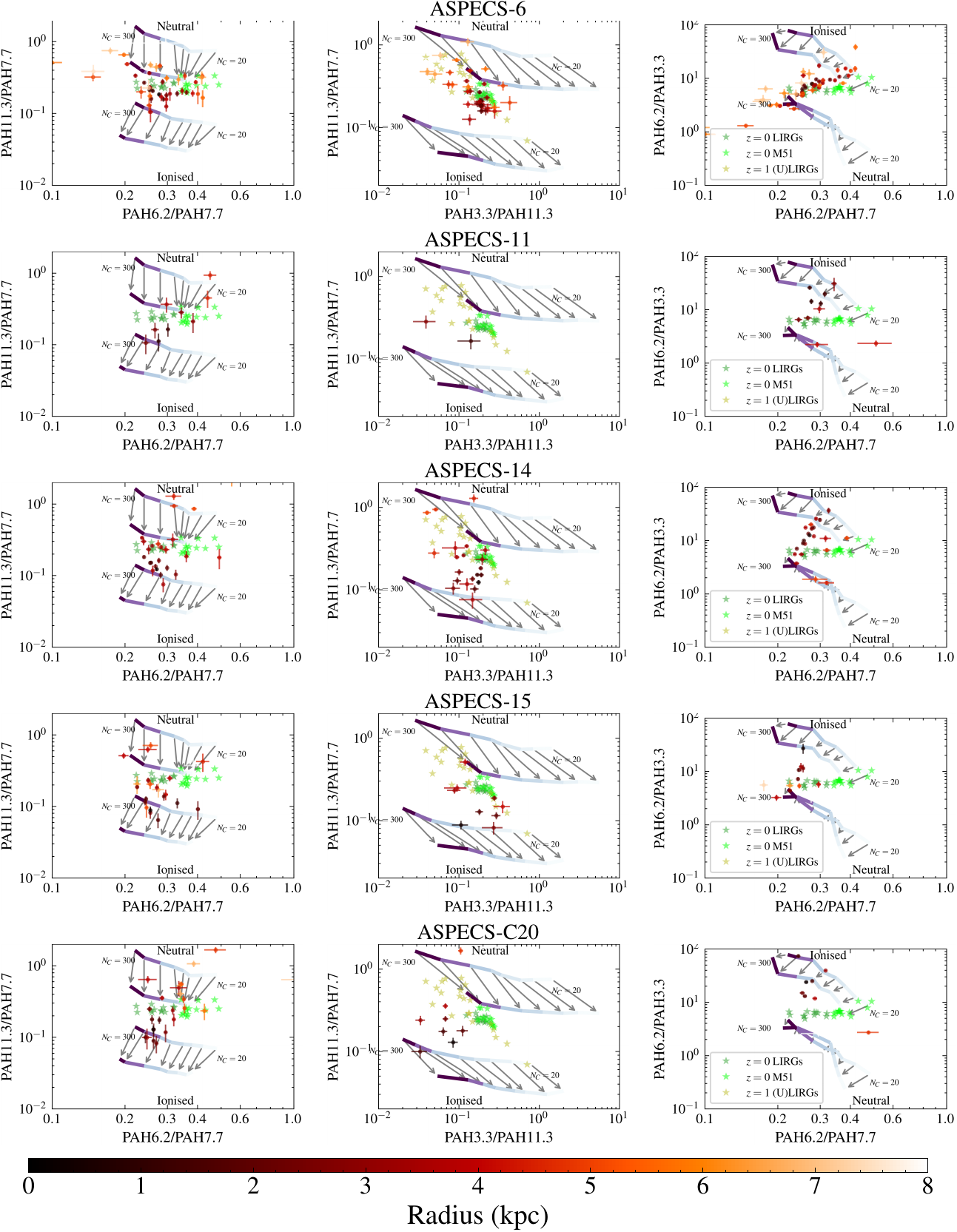}
    \caption{Plots of various PAH band ratios for the individual spaxels of the five ASPECS galaxies. We show the 11.3/7.7 vs 6.2/7.7 ratios in the first panel and the 11.3/7.7 vs 3.3/11.3 ratios in the second panel and the 6.2/3.3 vs 6.2/7.7 ratios in the third panel. We only show spaxels where there is at least a 3$\sigma$ measurement for both PAH ratios in each plot. The color of the points indicates the distance from the center of the galaxy in kpc. We overlay theoretical PAH grids for fully neutral and fully ionized PAHs, where each grid shows a variety of sizes from 20 carbon atoms to 300 carbon atoms at the local interstellar radiation field (ISRF) strength and at 1000$\times$ISRF \citep[][]{Rigopoulou2021, Rigopoulou2024}. The grey arrows point from the ISRF track to the 1000$\times$ISRF track.
    As a comparison sample we show the values of 12 star-forming regions from local LIRGs in dark green \citep[][]{Rigopoulou2024}, and from M51 in light green (see Appendix \ref{sec:M51}). For the 11.3/7.7 vs 3.3/11.3 ratios we also show $z\sim1$ PAH ratios of (U)LIRGs from \citet{McKinney2025} with the olive stars. 
    }
    \label{fig:Ratios} 
\end{figure*}

We plot the ratios of the individual spaxels in Fig. \ref{fig:Ratios} for each of the galaxies in comparison to local PAH ratios from JWST spectroscopy. We use the star-forming regions of NGC 3256 and NGC 7469 from the GOALS ERS program (ERS 1328, P.I. L. Armus and A. Evans) with PAH ratios extracted using \textsc{SPIRIT} in \citet{Rigopoulou2024}. As we corrected our templates for missing flux that would be included in the wings of Drude profiles, these ratios are directly comparable.

As NGC 3256 and NGC 7469 are Luminous Infrared Galaxies, we also compare to a more ``normal'' star-forming galaxy, M51, which has been observed with both NIRSpec and MIRI MRS from The JWST Whirlpool Galaxy Treasury (GO 3435, P.I. K. Sandstrom and D. Dale). We extracted spectra from 15 regions of M51, including isolated HII regions, which we fit with \textsc{SPIRIT} from \citet{Donnan24a}. This fitting is shown in Appendix~\ref{sec:M51}. We find the M51 PAH ratios to be largely consistent with the LIRGs but with slightly higher 6.2/7.7 PAH ratios as shown in Fig. \ref{fig:Ratios}. It is worth noting that the PAH ratios measured in local galaxies are on smaller spatial scales than the ASPECS galaxies which can explain the scatter between the local galaxy values which would become averaged out in observations of more distant galaxies.

We compare the inferred PAH ratios to theoretical PAH grids, calculated from Density Functional Theory of a collection of PAH molecules \citep[][]{Rigopoulou2021, Rigopoulou2024}, which are shown as grids in Fig. \ref{fig:Ratios}. These grids are produced for a fully ionized mix of PAHs and a fully neutral mix for a range of sizes from 20 carbon atoms to 300 carbon atoms. For each grid, the molecules are illuminated by an interstellar radiation field (ISRF) similar to the local solar neighbourhood, with an average photon energy of 12.4 eV, as well as 1000$\times$ the intensity.

We also show the radial distance of each spaxel in Fig. \ref{fig:Ratios} to further identify any radial trends, where the radius is measured as the projected distance from the center of the galaxy as measured from the F770W MIRI image. We note that this is a projected radius on the sky and thus is not corrected for orientation parameters if one assumes a disk. For ASPECS-15, as it is edge-on, there will be contribution from multiple radii along a given sight-line.

\section{Discussion}
\label{sec:Discuss}


\subsection{Radial Profiles}
PAH ratios have been measured extensively in the local universe with JWST, both with spectroscopy \citep[e.g.][]{Rigopoulou2024, Garcia-Bernete2022c, Garcia-Bernete2024b, Zhang2024, Lai2023, Lai2025} and photometry, particularly from the PHANGS sample \citep[e.g.][]{Baron2025, Whitcomb2025, Koziol2026}. From the PHANGS sample, a clear variation of the 3.3 PAH is seen as a function of radius, enabled by the significantly larger field of view afforded by photometry, where the 3.3 PAH strength increases relative to 11.3 with increasing radius \citep[][]{Sandstrom2023, Baron2025, Koziol2026}. The 11.3/7.7 PAH ratio also decreases as the 3.3/11.3 increases in local galaxies, suggesting PAHs become smaller and more ionized with radius \citep[][]{Koziol2026}. The opposite is true for local LIRGs however, where the 11.3 $\mu$m PAH appears more extended than the 6.2 and 7.7 $\mu$m PAHs \citep[][]{Diaz-Santos2011}, suggesting more neutral PAHs at larger radius.

The increase in 3.3 $\mu$m PAH correlates strongly with metallicity, where in extreme metal poor galaxies such as blue compact dwarf galaxies (BCD), the 3.3 $\mu$m PAH is significantly brighter than any other PAH feature \citep[][]{Lai2025, Tarantino2025}. The presence of smaller PAHs in low metallicity environments has been attributed to inhibited growth, where a lack of available carbon atoms from the gas phase results in the reduced of growth PAHs in the ISM and thus the PAH population has a smaller size distribution \citep[e.g.][]{Sandstrom2012, Whitcomb2024}. However, unlike the metallicity trends in the PHANGS sample, PAHs also tend to be more neutral at low metalicity in BCDs \citep[e.g.][]{Lai2025, delValle-Espinosa2026}. There are a few possible reasons for this. Firstly, small PAHs tend to be more neutral based on ionization balance calculations \citep[e.g.][]{Tielens2005} and secondly, photo-destruction of the weaker ionized PAHs, potentially due to a lack of shielding by molecular gas and dust, can lead to a more neutral PAH population.

\begin{figure*}
    \centering
        \includegraphics[width=\textwidth]{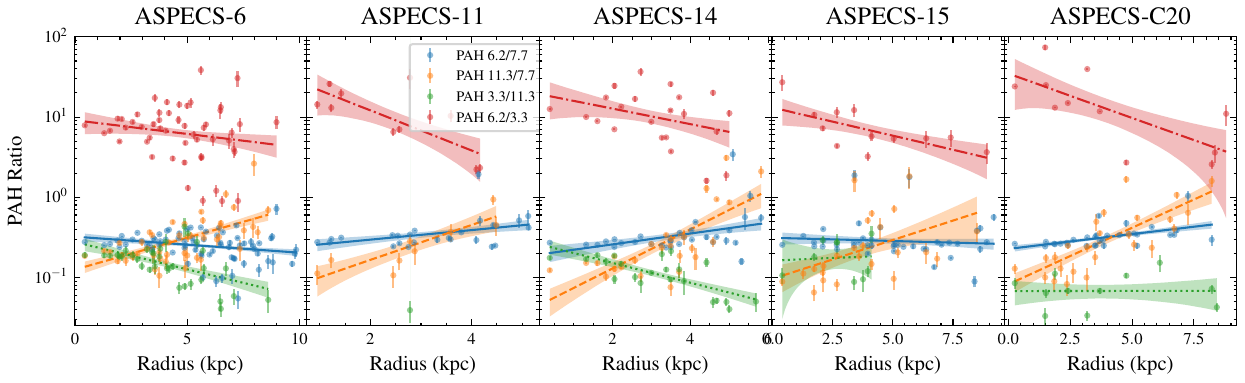}
    \caption{PAH ratios against radius (projected) for each of the galaxies. We plot the 6.2/7.7, 11.3/7.7, 3.3/11.3 and 6.2/3.3 PAH ratios. We show linear fits where there is a $>1\sigma$ measurement of the gradient i.e. a linear trend is present. The shaded regions show the 1$\sigma$ error. We also include a parameter for the intrinsic scatter in x and y, which produces a larger 1$\sigma$ region than is shaded.
    }
    \label{fig:RadProfiles} 
\end{figure*}

We show the radial profiles (radius as measured as distance from the nucleus) for our $z\sim1.1$ galaxies in Fig. \ref{fig:RadProfiles} where we have performed a power law fit, accounting for errors in both x and y axes, to check for a trend. We performed the fitting using MCMC sampling from \textsc{emcee}, where we include the errors in x and y in the variance of the log-likelihood. We additionally include the intrinsic scatter when calculating the total variance term.

It is worth noting that the points are not completely independent from each other as the regularisation/smoothing during the fitting requires neighbouring pixels not to vary too much from each other. Each pixel is also smaller than the PSF scale, particularly at 11.3 $\mu$m, due to the PSF convolution step in the modeling. At the longest wavelengths of MIRI, the PSF has a FWHM$\sim0.9''$ while each pixel in our PAH maps have a size of $0.2''$ (see section \ref{sec:Methods}) meaning that at most there are 5 pixels within the core of the PSF. 
We also include an intrinsic scatter parameter, where we find the intrinsic scatter to be larger than the measurement errors.

We see trends in the 6.2/3.3, 11.3/7.7 and 3.3/11.3 PAH ratios as a function of distance to the galaxy nuclei, particularly for ASPECS-6. The clearest trend is in the 6.2/3.3, for which there is a clear decrease with radius while there is an increase in the 11.3/7.7 PAH ratios with radius. As these two ratios are mostly sensitive to charge rather than size \citep[e.g.][]{Draine21, Rigopoulou2021}, these trends imply a greater fraction of neutral PAHs at larger radii. In ASPECS-6 and ASPECS-14, the 3.3/11.3 PAH ratio decreases with radius suggesting larger PAHs at larger radii. The 6.2/7.7 PAH ratio shows a much weaker trend, appearing to decrease with radius in ASPECS-6 but shows a gentle increase with radius in the other galaxies. As we discuss in section \ref{sec:Ancillary}, the behaviour of 3.3/11.3 and 6.2/7.7 PAH ratios with radius may be driven by differences in the hardness of the radiation field, especially considering that the PAHSPECS galaxies show PAH ratios more similar to the local LIRGs than M51 in Fig. \ref{fig:Ratios}.

Compared to the local galaxies in the PHANGS sample, we find the opposite trend with radius, where the 3.3 PAH emission is compact, resulting in a decreasing 3.3/11.3 PAH ratio with radius while  the 11.3/7.7 increases with radius suggesting PAHs become larger and more neutral in the outer regions of the galaxy. The anti-correlation of the 3.3/11.3 and 11.3/7.7 ratios is consistent with local galaxies but follows the opposite trend with radius \citep[][]{Baron2025, Whitcomb2025, Koziol2026}. If metallicity-driven effects are the primary driver of PAH band ratios, this might suggest an inversion of the typical metallicity gradients seen in the local universe, where instead of the metallicity decreasing with radius, it instead increases. Indeed, metallicity gradients are much more mixed at cosmic noon compared to the local universe, where most are flat and some actually show inverted gradients \citep[e.g.][]{Curti2020}.  

The trends we see in the ASPECS galaxies are however more consistent with local LIRGs rather than the PHANGS sample, where the 11.3/7.7 increase with radius \citep[][]{Diaz-Santos2011}. This suggests that the physical processes within local LIRGs may be more similar to the ASPECS galaxies. The ASPECS galaxies are indeed LIRGs if classified purely by luminosity ($L_{\rm IR}>10^{11} L_{\odot}$), however local LIRGs are more starburst in nature while the ASPECS galaxies are main-sequence. \citet{Diaz-Santos2011} suggests that the 11.3 $\mu$m emission is more extended due to more evolved stars powering the feature, however as we discuss in section \ref{sec:Ancillary}, we find that the stellar age peaks in the nucleus and so evolved stars are likely not driving the 11.3/7.7 ratio.

As an alternative to metallicity driven effects, destruction may be dominant in these galaxies, as small and ionized PAHs are thought to be easier to destroy in hard radiation fields, resulting in a larger and more neutral PAH population \citep[e.g.][]{Leach1986, Voit1992, Micelotta2010, Zhen2015,  Lange2025}. This is indeed the case in AGN \citep[e.g.][]{Stierwalt2013, Stierwalt2014, Garcia-Bernete2024b, Zhang2024}, where PAHs are particularly neutral. Considering the similarity in the radial trend to the local LIRGs, increased destruction is a plausible mechanism. To test these possible mechanisms, namely inhibited growth and destruction, we use the plethora of ancillary photometry from rest-frame UV to near-IR in section \ref{sec:Ancillary}.

One may suspect the lack of 3.3 $\mu$m PAH is a result of the relative low signal to noise of the short wavelength region of the spectrum, however similarly compact 3.3 PAH emission is also seen in MIRI LRS spectra of $z\sim1$ galaxies \citep[][]{Stone2026}, where maps have been created using slit stepping, which has a much higher signal to noise at the 3.3 PAH feature compared to our MRS data. Compared to the 6.2 $\mu$m PAH, the 3.3 $\mu$m PAH emission is more compact, consistent with our results.


\subsection{What is driving the observed PAH ratios?}
\label{sec:Ancillary}
To understand the physical conditions behind the observed PAH ratios, we use the extensive ancillary photometric data from HST \citep[][]{Illingworth2013} and JWST/NIRCAM \citep[][]{Rieke2023}, where SED fitting can provide measurements of stellar age, SFR, metallicity etc. The high spatial resolution of this data allows us to fit the photometry on a spatially resolved scale to produce maps of SFR, metallicity, $A_V$ for all our targets. Before fitting we convolved the higher resolution images to the scale of the lowest resolution filter (JWST/F480M) before reprojecting onto the pixel scale of our PAH maps (MIRI/MRS Ch3, 0.2$''$).

We use \textsc{BAGPIPES} \citep[][]{Carnall2018} to fit the SEDs of each spaxel. The model consists of a \citet{Bruzual2003} stellar population with a decayed exponential star-formation history, attenuation using the \citet{Calzetti2000} dust law and nebular emission from \textsc{Cloudy} \citep[][]{Ferland2017}. We fix the strength of the nebular emission as we find it is difficult to constrain with our photometric coverage. We use the value of $\log U$ from the gas component of the galaxy integrated values from \citet{Shivaei2024} with values of $\log U$ = [-2.8, -2.8, -2.9, -3.35, -2.91] for ASPECS-6, ASPECS-11, ASPECS-14, ASPECS-15, ASPECS-C20 respectively. We additionally place Gaussian priors on the stellar age and metallicity using the galaxy integrated values from \citet{Shivaei2024}. The fitting is performed using nested sampling to produce posterior probability distributions for all the parameters of interest and thus accurate error estimates. 

The maps of various properties inferred from the spatially resolved SED fitting are shown in Fig. \ref{fig:Pipes}. More detailed spatially resolved SED modeling for the ASPECS galaxies, including the MIRI photometry, can be found in \citet{Liao2026}. We find our results are consistent with \citet{Liao2026}, where the radial trends of age, stellar mass, SFR and sSFR match ours, although their more sophisticated modeling results in lower uncertainties on the model parameters.

We also produce maps of the intrinsic UV flux and intrinsic optical flux for each of the galaxies using the HST/F435W (207 nm restframe) for the UV and JWST/F115W (550 nm restframe) for the optical map. This allows us to calculate a map of the hardness of the radiation field as the ratio of UV/Optical flux and therefore investigate whether destruction of PAHs are driving the observed trends. We use the original filters to measure the hardness (only relying on the $A_V$ inferred from the SED modeling to correct the photometric flux) rather than integrating the best fit galaxy model throughout the entire UV band as extrapolating to the far-UV may not be reliable.

The hardness of the radiation field and specific star-formation rate appear to increase with radius, particularly for ASPECS-6. This may suggest increased photo-destruction of small and ionized PAHs at large radius is responsible for the larger and more neutral PAH population, rather than increased growth leading to larger PAH sizes. If increased growth were responsible we might expect an inversion of the expected metallicity gradients, where lower metallicity at low radius would reduce the carbon budget and lower the efficiency of grain growth preventing larger PAHs growing in the ISM \citep[e.g.][]{Tielens2008}. However, in lieu of gas-phase metallicity measurements, the decrease in stellar mass density with radius may suggest a typical metallicity gradient, assuming the mass-metallicity relation \citep[e.g.][]{Tremonti2004, Sanchez2013, Barrera-Ballesteros2016}, where mass has built up in the nucleus first and thus there is a lower metallicity at large radius. 

\begin{figure}
        \includegraphics[width=\columnwidth]{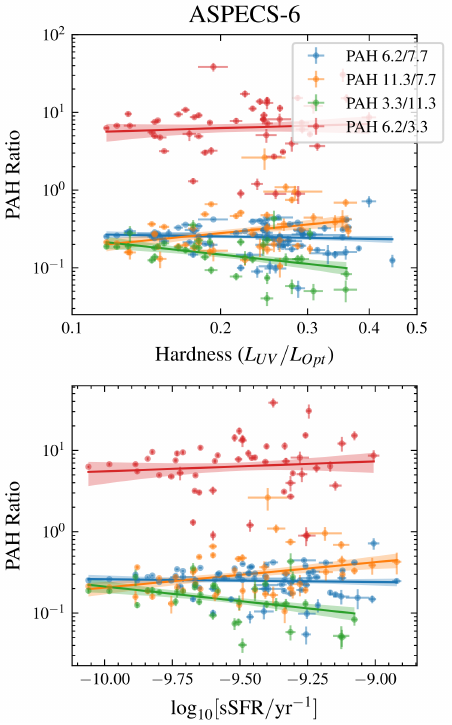}
    \caption{PAH ratios against the hardness of the radiation field (upper) as measured by the ratio of flux at 207 nm to 550 nm restframe and the specific star-formation rate (lower) for ASPECS-6. We plot the 6.2/7.7, 11.3/7.7, 3.3/11.3 and 6.2/3.3 PAH ratios. We show linear fits where there is a $>1\sigma$ measurement of the gradient i.e. a linear trend is present. The shaded regions show the 1$\sigma$ error.  We also include a parameter for the intrinsic scatter in x and y, which produces a larger 1$\sigma$ region than is shaded.
     }
    \label{fig:Hardness} 
\end{figure}

Therefore, considering the increased hardness at large radii, photo-destruction of the small and ionized PAHs, as they are more susceptible to destruction than large and neutral PAHs \citep[e.g.][]{Leach1986, Voit1992, Lebouteiller2007, O'Dowd2009, Micelotta2010}, is a more likely explanation. We plot the measured PAH ratios explicitly against the hardness of the radiation field in Fig. \ref{fig:Hardness} for ASEPCS-6. The remaining galaxies are shown in Fig. \ref{fig:Hardness2} however the fewer quality data points makes any possible trends harder to observe. For ASPECS-6, the 3.3/11.3, 6.2/3.3 and 11.3/7.7 PAH ratios show a trend with hardness and sSFR, where indeed, harder radiation fields show larger and more neutral PAHs. The 6.2/7.7 ratio is not consistent however, showing a slight increase in hard radiation fields. This trend is particularly driven by two regions of high 6.2/7.7 to the north west and south west of the galaxy.

While we have stated that harder radiation fields can cause increased destruction of the weakest PAH molecules, we also have to consider that a harder radiation field also alters the excitation of the surviving PAH molecules by altering the temperature distribution of the PAHs \citep[e.g.][]{Draine21}. The PAH ratio model grids from \citet{Rigopoulou2021} and shown in  Fig. \ref{fig:Ratios} only vary intensity and not hardness. The PAH model grids presented in \citet{Draine21} test the effect of excitation by different radiation fields, ranging from a young (3 Myr) starburst model from \citet{Bruzual2003} to the bulge of M31. \citet{Draine21} shows that the 11.3/7.7 PAH ratio is relatively insensitive to hardness, however the 6.2/7.7 and 3.3/11.3 PAH ratios increase in harder radiation fields, with the 6.2/7.7 increasing by a maximum factor of $\sim1.5$ and the 3.3/11.3 increasing by a maximum factor of $\sim2.5$.

Our hardness measurements of $\frac{F_{207\textrm{nm}}}{F_{550\textrm{nm}}}\sim0.1-0.5$, correspond to a modest range of the energy absorbed per photon of $\sim4 - 6$eV based on the \citet{Draine21} models. The full range of the \citet{Draine21} models grids span a range of $\sim2.7-8.6$eV and so we would not expect an increase in the 6.2/7.7 as large as $\sim1.5$ and thus hardness cannot fully explain why the 6.2/7.7 is higher in those two particular regions of ASPECS-6.

Considering that the 3.3/11.3 and 6.2/7.7 decrease in harder radiation fields in ASPECS-6 suggests that size, and in particular photo-destruction of small PAHs, is driving these particular ratios. Future work is needed to study ASPECS-6 in more detail, utilizing MUSE and ALMA data to understand the physical conditions behind the two regions of high 6.2/7.7.


\subsection{Weak 3.3 PAH at cosmic noon?}
In addition to the radial profiles of the 3.3/11.3 PAH ratios behaving differently compared to local galaxies, in some of our galaxies the overall values of the 3.3/11.3 PAH appear lower, with the 3.3 PAH feature barely detected in ASPECS-11, ASPECS-15 and ASPECS-C20. 

By pairing MIRI LRS spectra with Spitzer IRS spectra, \citet{McKinney2025} measured the 3.3/11.3 PAH ratios for a sample of massive galaxies at cosmic noon, $z\sim0.6-2.5$. The authors find that indeed the 3.3 PAH is relatively weaker than in local dusty LIRGs which we have plotted in Fig. \ref{fig:Ratios}. Compared to the PAHSPECS sample, these are more luminous, occupying more of the starburst region of the main-sequence while our sample are more typical but fainter.

\citet{McKinney2025} suggests that the increased fraction of large PAHs is due to coagulation, where in high-density environments, small PAHs coagulate to form larger PAHs, shifting the emission to longer wavelengths. This is indeed plausible in large and luminous galaxies where gas densities are thought to be high consistent with their starburst nature and supported by the detection of H$_2$O ice at $\sim3\mu$m.

In our sample the 3.3/11.3 PAH ratio decreases in the outer regions where the SFR is lower but the sSFR is higher. The coagulation scenario would require therefore higher densities at large radii. Assuming the Kennicutt-Schmidt relation \citep[][]{Kennicutt1998}, the gas surface density would be lower at high radii due to the low SFR, however a high gas surface density in the nucleus could be driven purely by the number density of clouds rather than the density of the individual molecular clouds. It is therefore plausible that individual molecular clouds are denser in regions of higher sSFR, where there is a lot of active star formation, but the SFR is low. This has been found in local galaxies where the surface brightness of individual clouds on small scales ($\sim10$s pc) does not track the $\sim$ kpc averaged gas surface density \citep[][]{Kim2026, Hughes2013, Leroy2013}.

An alternative scenario, rather than increased coagulation, is photo-destruction, which will preferentially destroy small PAHs resulting in the weak 3.3 $\mu$m PAH. Considering we also see more neutral PAHs at high radii, preferential destruction, of which small and ionized PAHs are more susceptible to, explains both the charge and size distribution.

\section{Conclusions}
We have presented the first spatially resolved PAH maps through MIRI/MRS spectroscopy at cosmic noon. Through careful modeling of the data cubes, including PSF convolution we have produced maps of the 6.2/7.7, 3.3/11.3, 11.3/7.7 and 6.2/3.3 PAH ratios. Our main findings are 
\begin{itemize}
    \item We find that the 3.3/11.3, 6.2/7.7, 6.2/3.3 PAH ratios decrease with radius while the 11.3/7.7 increases with radius. This is consistent with more neutral and large PAHs at higher radii. This is the opposite trend of local main-sequence galaxies but matches that of local LIRGs.
\end{itemize}

\begin{itemize}
    \item We find that the increase in neutral and large PAHs correlates with an increased hardness of the radiation field obtained from photometry, suggesting photo-destruction of ionized and small PAHs at higher radii rather than inhibited growth.
\end{itemize}

\begin{itemize}
    \item We find the dust continuum emission is compact compared to the PAH emission, consistent with local galaxies. In ASPECS-6 the continuum is elongated along the minor axis, possibly forming a bar structure. 
\end{itemize}

\begin{itemize}
    \item In ASPECS-6, we find two distinct regions of high 6.2/7.7, inconsistent with the 3.3/11.3 if purely size is driving the observed ratios. These regions also exhibit hard radiation fields, which may suggest an altered excitation.
\end{itemize}

\begin{itemize}
    \item We find that the 3.3 $\mu$m PAH emission is on average weaker relative to the 11.3 $\mu$m PAH than local galaxies, consistent with higher luminosity galaxies observed with MIRI/LRS. This is consistent with increased destruction of small PAHs at higher cosmic noon but potentially also due to increased coagulation of PAHs in denser molecular clouds.
\end{itemize}

These results suggest that the ISM conditions are different at cosmic noon, resulting in a different spatial distribution of PAH properties and thus some evolution from cosmic noon to the local universe. Follow-up detailed analysis of ASPECS-6 will lead to further understanding of the conditions at cosmic noon \citep[][]{Donnan2026c}. 
\newlength{\twocolwidth}
\setlength{\twocolwidth}{\columnwidth}
\begin{acknowledgments}
FRD thanks Francisco Rodr\'iguez Montero for the helpful discussion. FRD and KS acknowledge funding support from grant JWST-GO-05279.002. IS acknowledges fundings from the European Research Council (ERC) DistantDust (Grant No.101117541) and the Atracc\'{i}on de Talento Grant No.2022-T1/TIC-20472 of the Comunidad de Madrid, Spain.
L.A.B. acknowledges support from the Dutch Research Council (NWO) under grant VI.Veni.242.055 (\url{https://doi.org/10.61686/LAJVP77714}).
CML acknowledges the research project was supported by the Hellenic
Foundation for Research and Innovation (HFRI) under the ``2nd Call for
HFRI Research Projects to support Faculty Members $\&$ Researchers''
(Project Number: 03382).
MA is supported by FONDECYT grant number 1252054, and gratefully acknowledges support from ANID Basal Project FB210003, ANID MILENIO NCN2024\_112 and ANID + Vinculaci\'on Internacional + FOVI250261. 
RD acknowledges support from the INAF RF 2024 mini-grant  ``The interstellar medium at high redshift'' and from the PRIN MUR ``2022935STW'', RFF M4.C2.1.1, CUP J53D23001570006 and C53D23000950006.
This work is based on observations made with the NASA/ESA/CSA James Webb Space Telescope. The data were obtained from the Mikulski Archive for Space Telescopes at the Space Telescope Science Institute, which is operated by the Association of Universities for Research in Astronomy, Inc., under NASA contract NAS 5-03127 for JWST. These observations are associated with program \#5279. The data can be accessed at \url{https://doi.org/10.17909/t4rj-fm69}.
Support for program \#5279 was provided by NASA through a grant from the Space Telescope Science Institute, which is operated by the Association of Universities for Research in Astronomy, Inc., under NASA contract NAS 5-03127.
We acknowledge the JADES team for their public data products available at \url{https://jades-survey.github.io}. We thank the referee for their careful and thorough review of the paper and their feedback.
\end{acknowledgments}

\appendix
\section{Strength of Regularisation}
\label{sec:Reg}
The level of smoothing/spatial correlation that is preferred by the fit depends on the choice of the factor, $\Gamma$ in equation \ref{eqn:prob}, which controls the strength of the regularisation and thus penalty factor that is applied to the $\chi^2$ for non-smooth solutions. The appropriate value of $\Gamma$ to use is not well defined and therefore we test a variety of values with ASPECS-6 to check how the result depends on the choice of $\Gamma$. If $\Gamma$ is too low, the inferred maps will be overly noisy while if $\Gamma$ is too high, the maps will be too smoothed and therefore one looses any intrinsic structure and may overestimate the size of the region of the emission.

\begin{figure*}
        \includegraphics[width=\textwidth]{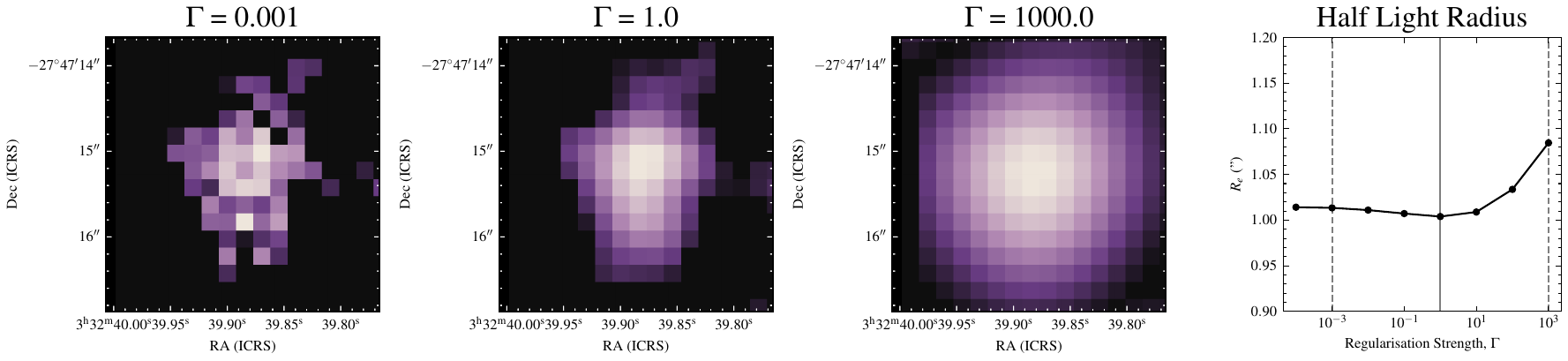}
    \caption{Demonstration of the effect of different values of the regularisation strength, $\Gamma$ in equation \ref{eqn:prob}, where the first three panels show the inferred PAH 6.2 maps for ASPECS-6. The first panel shows a low value of $\Gamma$ where the  map is overly noisy while the third panel is over-smoothed with $\Gamma$ set too high. The second panel is the optimum value of $\Gamma$. The final panel shows how the size as measured by the half light radius, changes with $\Gamma$, where at high values the size is over-predicted. The vertical lines display the values of $\Gamma$ in the left 3 panels.
    }
    \label{fig:GammaTest} 
\end{figure*}

We fit the data of each of the emission features with the following values of $\Gamma = 0.0001, 0.001, 0.01, 0.1, 1, 10, 100, 1000$. To judge the optimal value of $\Gamma$ we visually inspect the inferred maps as well as measure the half light radius. We calculate the half-light radius numerically by stepping out from the centre of the galaxy until half the flux is enclosed. 
We present 3 values of $\Gamma$ in Fig. \ref{fig:GammaTest} for the 6.2 $\mu$m PAH where the first case shows a noisy map where the strength of the regularisation is too low, the second case is the optimum value while in the final case the value of $\Gamma$ is too high. We show the measured half light radius for each value of $\Gamma$ in the final panel of Fig. \ref{fig:GammaTest}. We choose the optimum value of $\Gamma$ as that which gives a smooth solution while not overestimating the size of the emission. Eventually at very high values of $\Gamma$, the model will simply not fit the data well where the $\chi^2$ of the fit will increase as the over-smoothed emission (such as seen at $\Gamma=1000$ in Fig. \ref{fig:GammaTest}), once convolved with the PSF is simply not a good fit to the data.

We find that the optimum values are $\Gamma=1$ for the 6.2 $\mu$m and 7.7 $\mu$m PAH features, $\Gamma=0.01$ for the 11.3 $\mu$m PAH, $\Gamma = 10$ for the 3.3 $\mu$m PAH and $\Gamma = 100$ for [ArII] (6.98 $\mu$m). 

We also test whether the choice of regularisation affects the trends of the PAH ratios with radius as over-smoothing on of the PAH maps relative to another may lead to an artificial radial trend. This is a particular worry for the 11.3 $\mu$m PAH considering its location in channel 4 of MIRI at $\sim24 \mu$m and the subsequent large PSF. 

\begin{figure*}
        \includegraphics[width=\textwidth]{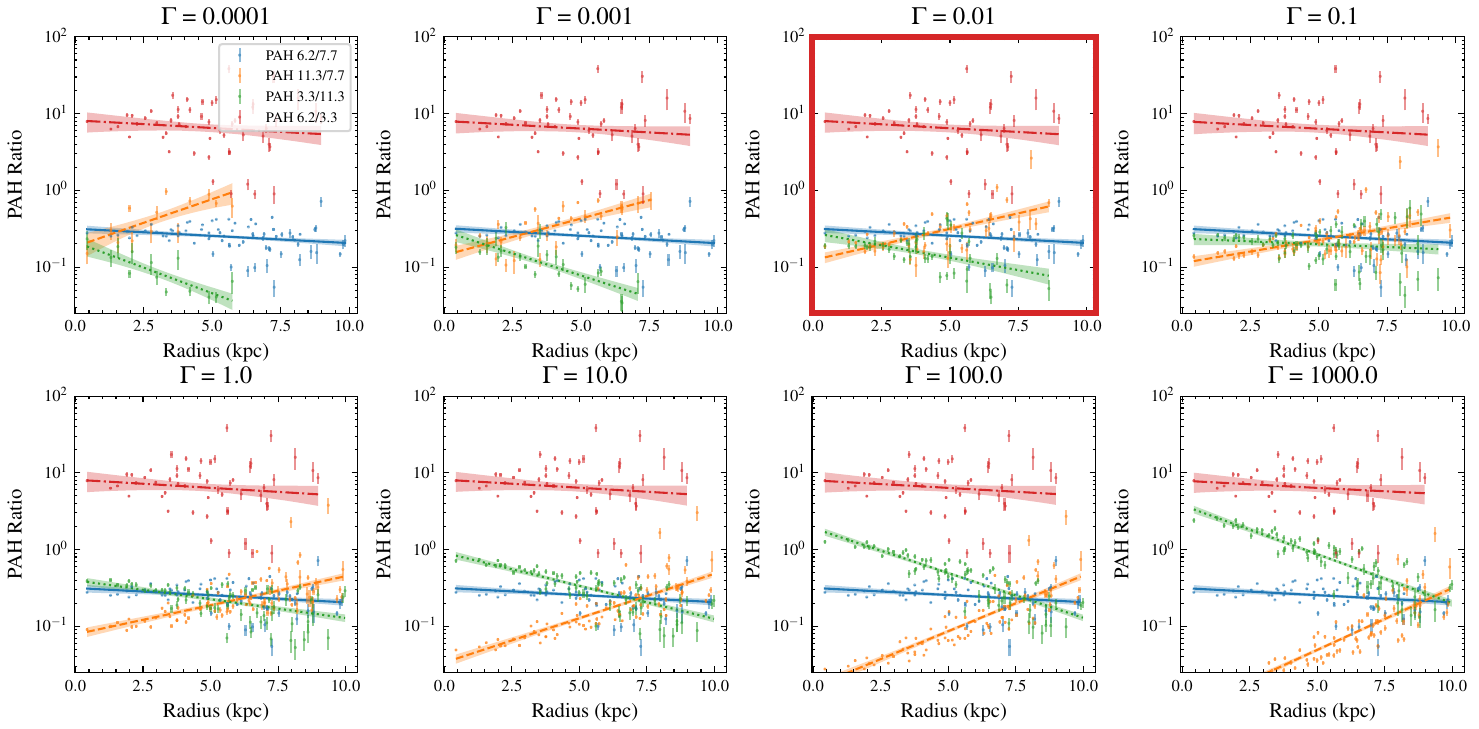}
    \caption{Testing how the strength of regularisation affects the observed radial trends in the PAH profiles. We show the radial profiles of the various PAH ratios for different strengths of the regularisation via the parameter $\Gamma$ for the fitting of the 11.3 $\mu$m PAH cube. The value of $\Gamma$ remains fixed at $\Gamma=1$ for the 6.2 $\mu$m and 7.7 $\mu$m PAH cubes, and $\Gamma = 10$ for the 3.3 $\mu$m PAH cube. The red box highlights the assumed value of $\Gamma = 0.01$ used through this work for the 11.3 $\mu$m PAH. Note only the 11.3/7.7 and 3.3/11.3 ratios change in each panel as only the regularisation strength for the 11.3 $\mu$m PAH is varied. This test shows that an artificial trend is induced at high values of $\Gamma$ as the size of the PAH emission region is overestimated, however the trend does remain at low $\Gamma$ proving there is a physical radial trend in the 11.3/7.7 and 3.3/11.3 PAH ratios.
    }
    \label{fig:GammaTest2} 
\end{figure*}

We perform this test on ASPECS-6 where we vary the value of $\Gamma$ in equation \ref{eqn:prob} for the fit to the cube containing the 11.3 $\mu$m PAH feature. The resulting radial profiles are shown in Fig. \ref{fig:GammaTest2}, where we show how all the PAH ratios change with radial distance as measured from the nucleus (the same as Fig. \ref{fig:RadProfiles}). It is worth noting that only the 3.3/11.3 and 11.3/7.7 PAH ratios vary between panels as the value of $\Gamma$ is changed for the 11.43 $\mu$m PAH fit. The red box highlights $\Gamma=0.01$ which was the chosen value for the analysis in this work from the previous test. 

\begin{figure*}
\centering
        \includegraphics[width=\twocolwidth]{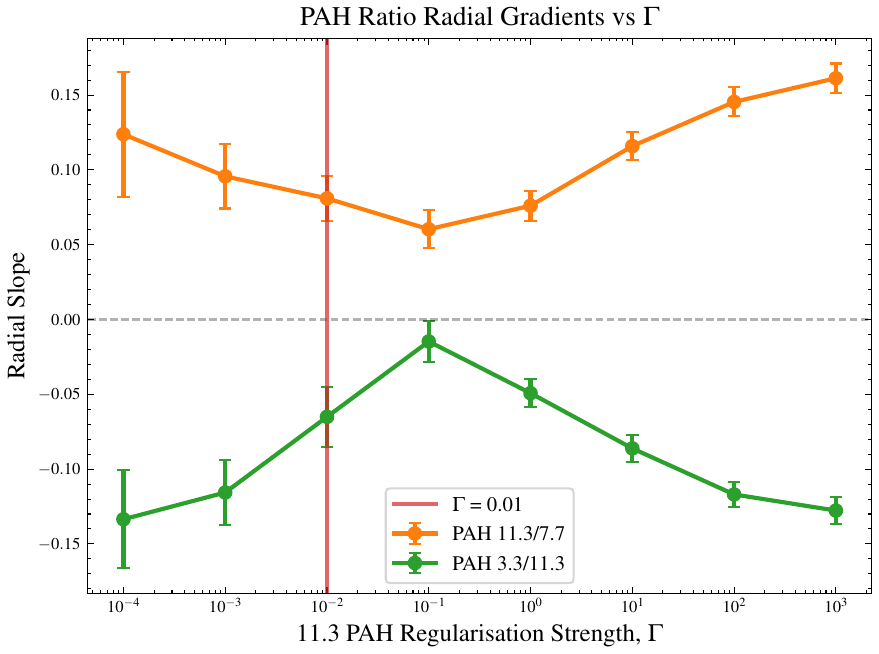}
    \caption{The measured gradient of the power law fits to the 11.3/7.7 and 3.3/11.3 PAH ratios as a function of the strength of regularisation via the parameter $\Gamma$ applied to the 11.3 $\mu$m PAH fits. The power law fits are shown in Fig. \ref{fig:GammaTest2}. The vertical red line shows the value assumed throughout this work where $\Gamma=0.01$. The
    }
    \label{fig:GammaTest3} 
\end{figure*}

The trends remain present for all value of the regularisation with the 11.3/7.7 showing a positive trend with radius while the 3.3/11.3 shows a negative trend. However the gradient does change as expected. We plot the measured gradient of the fits to each PAH ratio as a function of $\Gamma$ in Fig. \ref{fig:GammaTest3}. At high values of $\Gamma$, the 11.3 $\mu$ PAH map is over-smoothed by the regularisation and thus the size of the emission is over-predicted (as demonstrated in Fig. \ref{fig:GammaTest}) leading to an artificial steeper gradient. When $\Gamma$ is too low, the inferred 11.3 $\mu$m PAH map is noisier, leading to fewer spaxels reaching the $3\sigma$ threshold to be included in the radial trend plot in Fig. \ref{fig:GammaTest2}, particularly at higher radii. This means the measured gradient is from a line fitted to fewer points, restricted to the inner regions of the galaxy and thus a different gradient is measured (this can bee seen in the top left panels of \ref{fig:GammaTest2} where only spaxels out of $\sim 5$ kpc have a $>3\sigma$ measurement of the PAH ratio). This test demonstrates that our radial trends are indeed accurate and physical, where the 11.3/7.7 increases with radius while the 3.3/11.3 decrease with radius.

\section{M51 Fits}
\label{sec:M51}
\begin{figure*}
        \includegraphics[width=\textwidth]{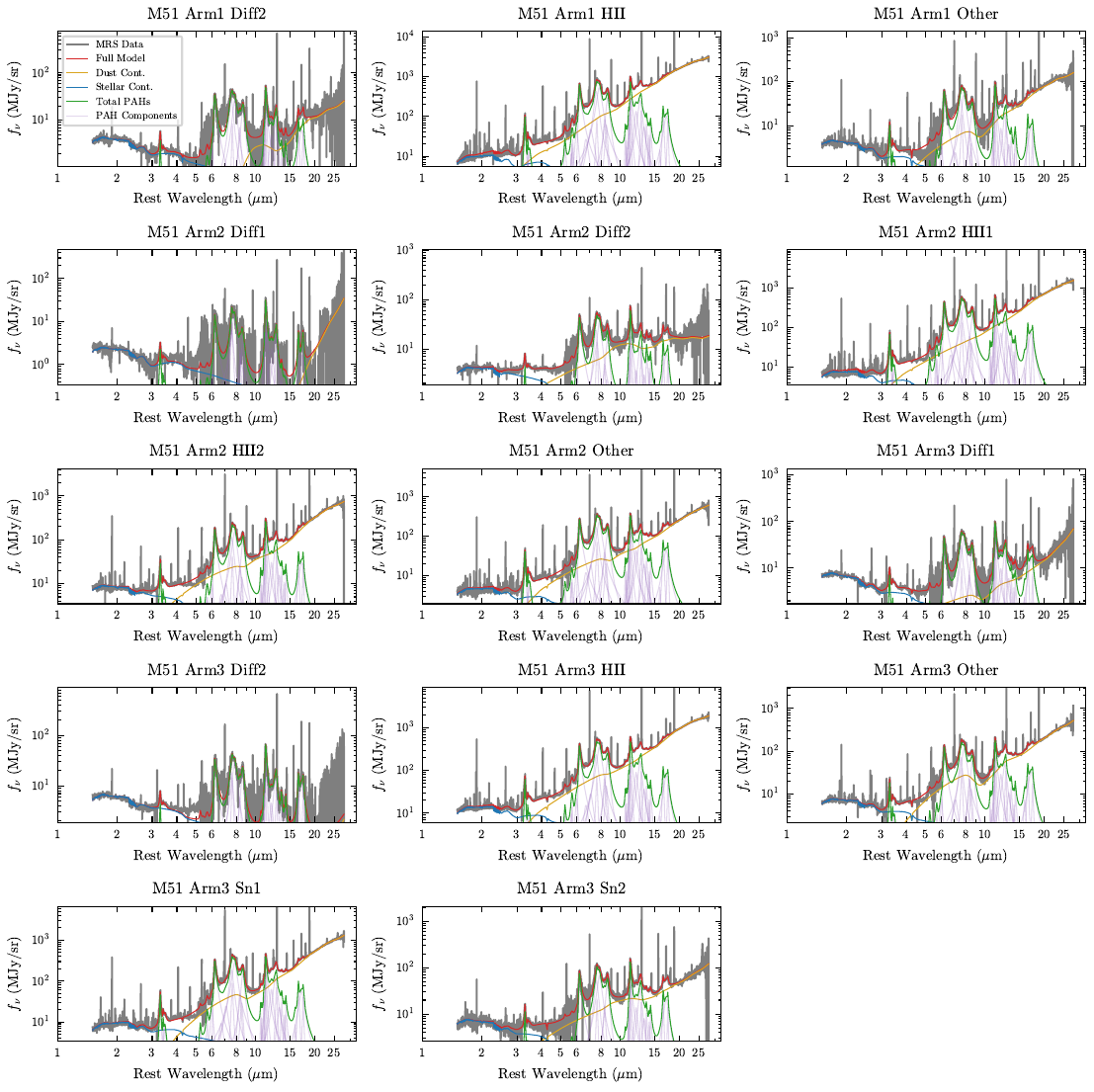}
    \caption{Fits to 14 spectra from different regions of M51 using \textsc{SPIRIT}, the spectral decomposition method presented in \citet{Donnan24a}, to infer PAH fluxes. The PAH ratios are shown as cyan stars in Fig. \ref{fig:Ratios}}
    \label{fig:M51Fits} 
\end{figure*}

To provide a comparison sample of typical star-formation in a local galaxy, in addition to the local LIRG sample, we fit the spectra extracted from 14 regions of M51 and fit the spectra with \textsc{SPIRIT}\footnote{\url{https://github.com/FergusDonnan/SPIRIT/}} (SPectral InfraRed Inference Tool ) which was presented in \citet{Donnan24a} to extract PAH fluxes. The regions span a range of environments including diffuse regions (Diff), HII regions (HII) and supernovae regions (Sn). These fits are shown in Fig. \ref{fig:M51Fits}. 

\section{\textsc{BAGPIPES} Fits}
\begin{figure*}
        \includegraphics[width=\textwidth]{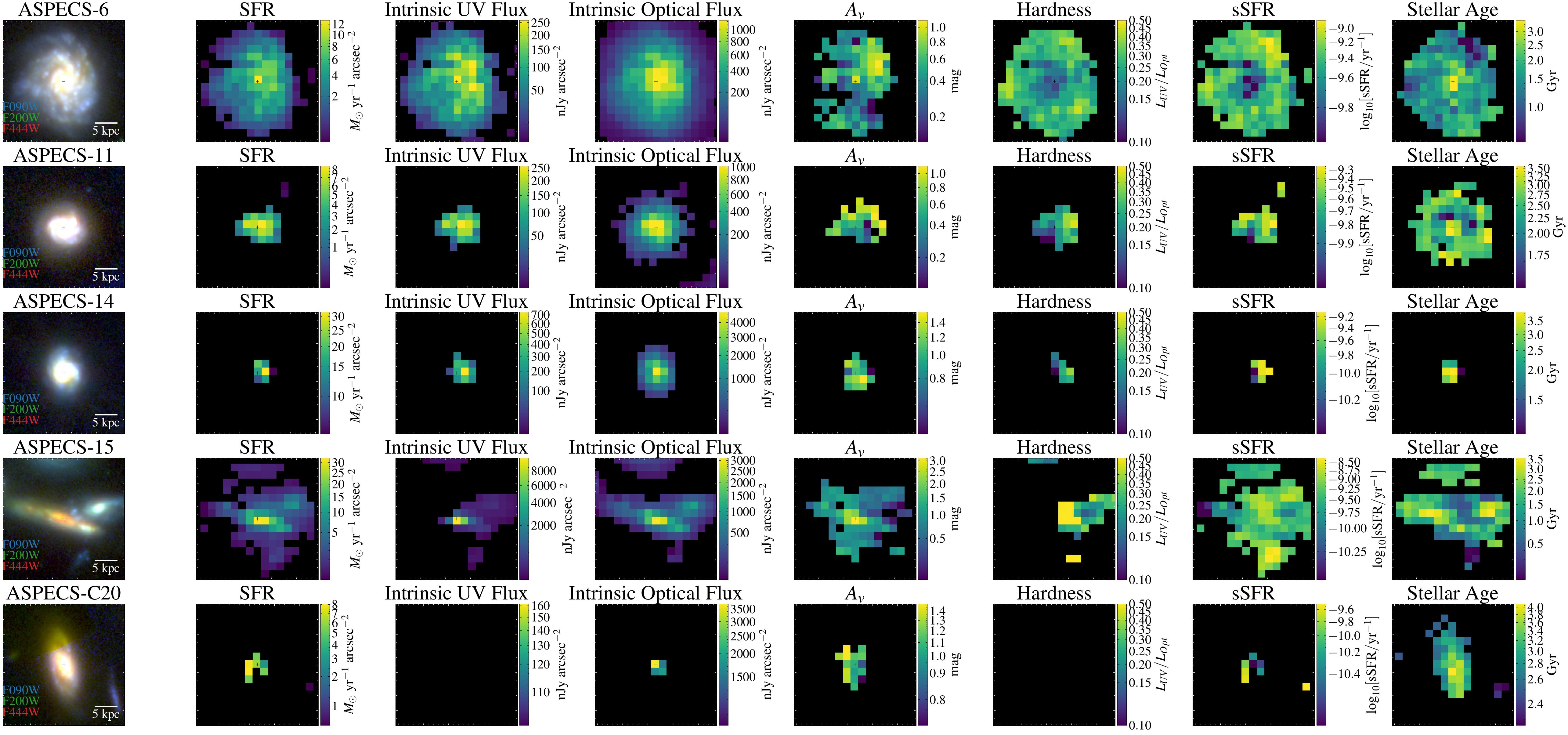}
    \caption{Maps of various galaxy properties extracted through spatially resolved SED fitting of HST and JWST/NIRCAM photometry with \textsc{BAGPIPES}. The left most column shows a three color JWST/NIRCAM image with the subsequent columns showing star-formation rate (SFR), dust corrected UV flux (207 nm), dust corrected optical flux (550 nm), the A$_v$, the hardness of the radiation field as measured by the ratio of UV/Optical flux (207 nm / 550 nm), the specific star-formation rate (sSFR) and stellar age. We only show spaxels where the property is measured to $>5\sigma$. }
    \label{fig:Pipes} 
\end{figure*}

We fit the photometry of the 5 galaxies in section \ref{sec:Ancillary} with \textsc{BAGPIPES} \citep[][]{Carnall2018}. The results of the fitting are shown in Fig. \ref{fig:Pipes}.

\begin{figure*}
        \includegraphics[width=\textwidth]{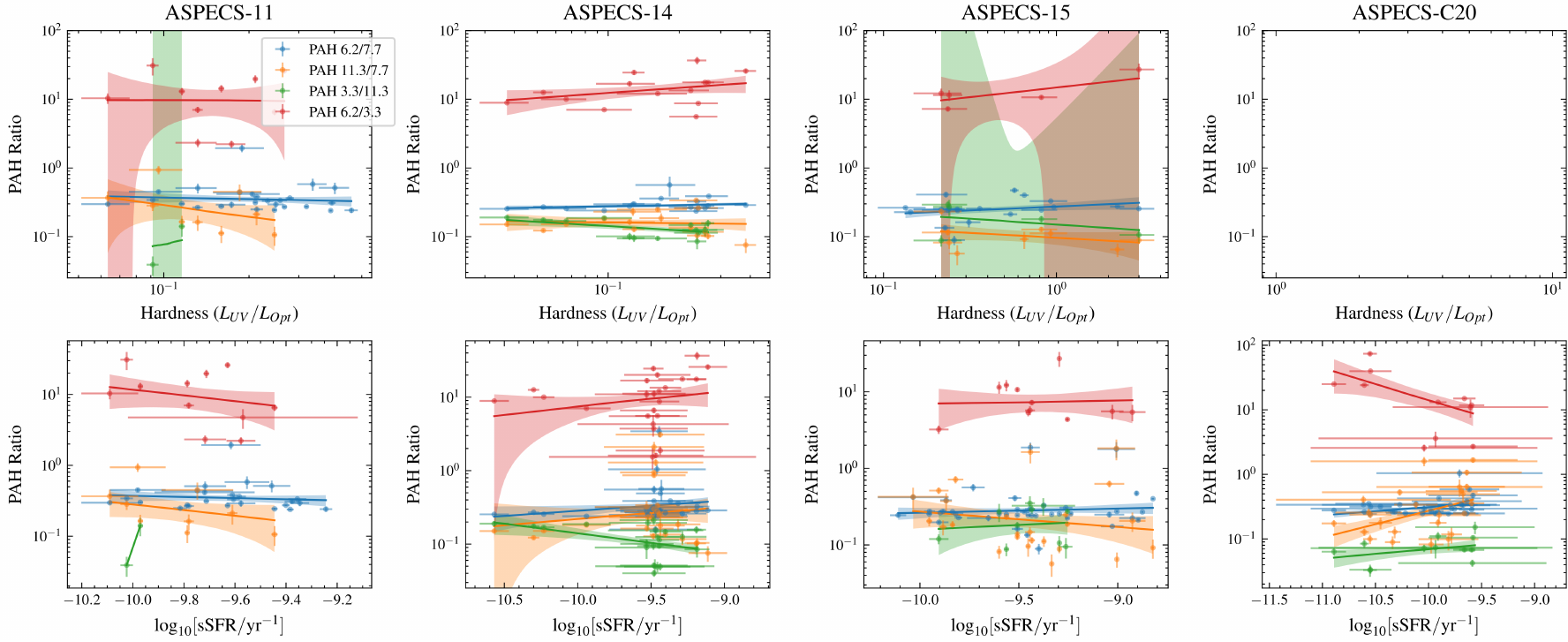}
    \caption{PAH ratios against the hardness of the radiation field (upper) and the specific star-formation rate (lower) for the remaining four galaxies. We plot the 6.2/7.7, 11.3/7.7, 3.3/11.3 and 6.2/3.3 PAH ratios. We show linear fits where there is a $>1\sigma$ measurement of the gradient i.e. a linear trend is present. The shaded regions show the 1$\sigma$ error.
     }
    \label{fig:Hardness2} 
\end{figure*}

We show plot the 6.2/7.7, 11.3/7.7, 3.3/11.3 and 6.2/3.3 PAH ratios against the hardness of the radiation field and specific star-formation rate for ASPECS-6 in Fig. \ref{fig:Hardness}. The remaining four galaxies are shown in Fig. \ref{fig:Hardness2}

\bibliography{sample701}{}
\bibliographystyle{aasjournalv7}



\end{document}